\documentclass[12pt]{article}

\usepackage[utf8]{inputenc}
\usepackage[T1]{fontenc}
\usepackage[english]{babel}
\usepackage{lmodern}

\usepackage{setspace}
\usepackage[margin=1in]{geometry}
\usepackage{microtype}

\usepackage{tabularx}
\usepackage{array}
\usepackage{booktabs}
\newcolumntype{L}[1]{>{\raggedright\arraybackslash}p{#1}}
\newcolumntype{Y}{>{\raggedright\arraybackslash}X}
\usepackage{graphicx}
\graphicspath{{figures/}}
\usepackage{float}
\usepackage[labelfont=bf]{caption}
\usepackage{enumitem}
\setlist{itemsep=0.5ex, topsep=0.5ex, parsep=0.5ex}
\usepackage{xcolor}
\definecolor{sectionblue}{RGB}{0,76,153}
\usepackage{sectsty}
\sectionfont{\color{sectionblue}\large\bfseries\sffamily}
\subsectionfont{\color{sectionblue}\normalsize\bfseries\sffamily}
\subsubsectionfont{\color{sectionblue}\normalsize\sffamily}

\usepackage{amsmath, amssymb, amsthm, bm}
\DeclareMathOperator{\logit}{logit}
\newtheorem{proposition}{Proposition}
\newcommand{\code}[1]{\texttt{\small #1}}

\usepackage[authoryear,round]{natbib}
\usepackage[colorlinks, citecolor=blue, urlcolor=blue, linkcolor=blue, bookmarks=true]{hyperref}

\title{Measuring Colombia's Informal Output:\\
Relative Earnings and Identification in a Bayesian Two-Sector Model, 1993--2026}

\author{
  Jaime A. Vera-Jaramillo\thanks{Email: \href{mailto:jaimevera1107@gmail.com}{jaimevera1107@gmail.com},
  \href{mailto:ja.veraj@uniandes.edu.co}{ja.veraj@uniandes.edu.co};
  ORCID: \href{https://orcid.org/0000-0001-7611-6251}{0000-0001-7611-6251}}
}
\date{November 2026}

\begin{document}
\maketitle

\begin{abstract}
How much of Colombia's measured output is produced by informal work? The answer cannot be read from the labor-informality rate: Colombia combines an informality rate above 50\% with a much smaller informal share of output, because informal work is less productive. We estimate a Bayesian two-sector business-cycle model on quarterly national accounts and on reconstructed employment and informality series for 1993--2026. Identification comes from a new quarterly series of informal-to-formal labor earnings built from household-survey microdata for 2007--2026. Formal production uses capital and labor, informal production is labor based, and an accounting relationship maps relative earnings, the informality rate and the formal capital share into the informal output share. We show that aggregate data alone cannot pin down the size of the informal sector, and that relative earnings identify its quarterly path: from 2007 the path is invariant, within 0.3 percentage points, to the model's steady state. The informal share declines from about 30\% of GDP in 1993 to 20\% in 2026Q2, while labor informality falls from about 69\% to 54\%. The estimates lie within the range of published shadow-economy series for Colombia, close to the enhanced MIMIC estimates of Asllani, Dell'Anno and Schneider for 2009--2020. Sensitivity analysis quantifies the role of the capital share and of non-wage labor costs.
\end{abstract}



\section{Introduction}
\label{sec:introduction}

An economy can have a high proportion of informal workers without producing an equally high proportion of its output informally. Colombia is a clear case. The national household-survey series used in this paper places labor informality at about 54\% of employment in 2026, but work in small, unregistered or own-account activities is much less productive than registered wage employment. Turning a headcount into a share of value added therefore requires information on that productivity gap, as well as a definition of what counts as informal production.

The indicators available in Colombia measure different things. Labor informality counts \emph{workers}. DANE's multidimensional business-informality indicator counts \emph{establishments}, and it classified 90.2\% of microbusinesses as informal in 2024 \citep{DANE_IMIE}. The statistical office's estimate of the non-observed economy, 22.4\% of gross production in the 2015-base accounts, covers informal, underground and illegal activities together with statistical adjustments \citep{DANE_SIECI}. This paper estimates something else, namely the share of \emph{measured value added produced by informal work} in an explicit two-sector production model. Because the units, denominators and coverage differ, none of these figures should be read as an estimate of the others.

Dynamic general equilibrium (DGE) models are a common tool for inferring the size of the informal or shadow economy. Calibrated versions fix its level at an external benchmark and model its dynamics from there \citep{ElginOztunali2012,SolisGarciaXie2017}, while Bayesian DSGE versions estimate latent sectoral activity jointly with the structural shocks \citep{OrsiRaggiTurino2014}. Either way, the level has to be identified, and aggregate data are often not enough to tell a large informal sector from a small one, because different shares can be reconciled with the same aggregates through changes in unobserved sectoral productivity and in the implied steady state. The contribution of this paper is to bring a direct measurement to bear on that problem.

The model is deliberately parsimonious. Competitive formal firms use capital and labor, and informal activity uses labor only, as in a familiar strand of the literature \citep{IhrigMoe2004,ElginOztunali2012}. A representative household divides its work between the two sectors, accumulates capital and chooses consumption, and aggregate shocks, a formality wedge and informal productivity generate fluctuations. We estimate the model by Bayesian methods with a Kalman-filter likelihood. The macroeconomic data are DANE's 2015-reference national accounts, with reconstructed quarterly observations before 2005, and total employment and labor informality come from the ICML series, which reconstructs the national labor market from 1993 \citep{VeraJaramillo2025ICML,VeraJaramillo2026Zenodo}.

The additional source of identification is a quarterly series of \emph{relative labor earnings} built from DANE's household-survey microdata since 2007. It distinguishes groups of workers and, on the formal side, is converted from reported wages to labor costs with the statutory non-wage costs of employment. In the model, the ratio of informal to formal output equals the ratio of workers times relative value added per worker, and relative value added per worker can be written in terms of the survey earnings ratio and the formal capital share. The result is a simple link between the data and the quantity of interest (Proposition~\ref{prop:share}), in which every assumption is explicit and can be changed.

The paper makes four contributions. It provides a quarterly estimate of informal output in Colombia from 1993 to 2026 that uses all the observed and reconstructed information available, together with the other series the model delivers, such as informal and formal GDP, relative productivity and the structural shocks. It shows, with likelihood profiles, prior--posterior comparisons and simulations with known parameters, that aggregate data alone do not determine the size of the informal sector, which helps explain why DGE models in the literature arrive at very different levels for the same country. It then shows that relative earnings resolve the problem for the object of interest: once they are observed, the estimated \emph{path} of informal output no longer depends on the steady-state share assumed by the model, and simulations recover the true share precisely. Finally, the data and code are public, so the series can be updated every quarter and re-estimated under other definitions, technologies and wage-cost conversions.

In the preferred specification, informal work accounts for 30.3\% of measured output in 1993 and 20.0\% in 2026Q2. The decline mirrors the fall of labor informality from about 69\% to 54\% of employment, and by the end of the sample informal value added per worker is about one-fifth of its formal counterpart. These estimates lie within the range of published shadow-economy estimates for Colombia, close to the enhanced MIMIC series in 2009--2020 and to the Medina--Schneider series in 2008--2015 \citep{SchneiderBuehnMontenegro2010,MedinaSchneider2019,AsllaniDellAnnoSchneider2024,ElginKoseOhnsorgeYu2021}, although those series measure related rather than identical concepts. Since this is a measurement exercise, the paper reports the level of informal output together with the evidence and the accounting assumptions behind it, so that each of them can be checked and improved.

Section~\ref{sec:background} reviews the measurement literature and Section~\ref{sec:data} describes the data and their provenance. Sections~\ref{sec:model} and~\ref{sec:estimation} present the model and the estimation strategy, Section~\ref{sec:identification} studies identification and Section~\ref{sec:results} reports the estimates. Section~\ref{sec:discussion} discusses how to interpret them, Section~\ref{sec:limitations} describes the extensions that would sharpen them and Section~\ref{sec:conclusions} concludes.

\section{Background and related literature}
\label{sec:background}

\subsection{Measuring the informal economy}

The size of the informal or shadow economy cannot be observed directly, and the literature has developed several indirect approaches \citep{SchneiderEnste2000, MedinaSchneider2018}. Currency-demand methods attribute part of the demand for cash to transactions that avoid taxation or registration. For Colombia, \citet{ArangoMisasLopez2006} estimate the underground economy from 1976 to 2003 with a cash-demand function in state-space form. Multiple-indicators multiple-causes (MIMIC) models treat the shadow economy as a latent variable linked to causes such as taxation and regulation and to indicators such as currency and labor participation \citep{MedinaSchneider2018}. Panels for Colombia and most other countries are available from \citet{SchneiderBuehnMontenegro2010}, \citet{MedinaSchneider2019} and, with an enhanced MIMIC, \citet{AsllaniDellAnnoSchneider2024}; \citet{Vuletin2008} estimates the informal economy of Latin America and the Caribbean with a structural equation model, and the World Bank database of \citet{ElginKoseOhnsorgeYu2021} collects DGE and MIMIC estimates. These methods produce a time path but need an external value to fix its level, and their structural interpretation is limited.

Model-based approaches use the equilibrium conditions of an economy with a formal and an informal sector. \citet{IhrigMoe2004} and \citet{BusatoChiarini2004} study the allocation of labor between a taxed formal sector and an untaxed informal one. \citet{ElginOztunali2012} calibrate a two-sector dynamic general equilibrium model to produce shadow-economy estimates for many countries, fixing the level in a base year with an external estimate. \citet{SolisGarciaXie2017} show that the growth rate imposed on shadow productivity in that approach is not consistent with the model's balanced-growth conditions and derive it from the observed trends instead; the level is again fixed with an external base-year value. \citet{OrsiRaggiTurino2014} estimate an underground economy for Italy in a Bayesian DSGE model with limited tax enforcement; observed consumption, investment, tax revenues and regular earnings identify the latent sector, and the prior is centered on the official estimate of the statistical office. They note that a key parameter of the underground technology has a posterior almost identical to its prior.

\subsection{Informality in business-cycle models}

A second literature uses informality to explain business-cycle facts in emerging economies. \citet{RestrepoEchavarria2014} shows that if national accounts measure informal output incompletely, measured consumption is more volatile than measured output. \citet{FernandezMeza2015} estimate a model of informal employment for Mexico and find that informality is countercyclical and amplifies the response to productivity shocks. The stochastic trend of the present model follows \citet{AguiarGopinath2007}, who show that permanent productivity shocks account for much of the cycle in emerging economies.

\subsection{Identification}

Identification in DSGE models has been studied in general by \citet{CanovaSala2009}, \citet{Iskrev2010} and \citet{KoopPesaranSmith2013}. Bayesian estimation always delivers a posterior, so a parameter that the likelihood does not inform is reported with a posterior close to its prior. The present paper applies three standard checks to the parameter that matters for the measurement of the informal economy: the ratio of posterior to prior dispersion, the profile likelihood, and the recovery of a known value in simulated data.

\subsection{Official statistics in Colombia}

DANE publishes quarterly national accounts in volumes chained to the 2015 base. Its national accounts include an imputation for the non-observed economy, estimated at 22.4\% of total production in the 2015 base \citep{DANE_SIECI}; this concept refers to gross production, not value added, and includes underground, illegal and informal activities and statistical corrections. Labor informality is measured by the household survey (GEIH) and has changed definition over time; the ICML project reconstructs consistent national and territorial series from 1993 \citep{VeraJaramillo2025ICML}.

\section{Data}
\label{sec:data}

\subsection{Sources and provenance}

Table~\ref{tab:data} lists the series. The national accounts are quarterly volumes in the 2015 base without seasonal adjustment. From 2005 on they are DANE's original series. For 1993--2004 they are retropolated: GDP, consumption and investment are disaggregated proportionally with the seasonal profiles of the official series, and government consumption with the Denton--Cholette method \citep{Denton1971}. Employment and the informality rate come from the GEIH universe of ICML release 2.6.1 \citep{VeraJaramillo2025ICML, VeraJaramillo2026Zenodo} and are averaged from monthly to quarterly values. Because the informality rate is informal over total employment, it is consistent with the employment series by construction.

\begin{table}[H]
\centering
\caption{Observables and provenance.}
\label{tab:data}
\footnotesize
\begin{tabularx}{\textwidth}{L{3.6cm} L{3.7cm} Y}
\toprule
\textbf{Series} & \textbf{Source} & \textbf{Provenance by period} \\
\midrule
GDP, private consumption, gross fixed capital formation, government consumption & DANE quarterly national accounts, base 2015, chained volumes & 2005--2026: original series. 1993--2004: retropolated quarterly series. \\
Total employment & ICML 2.6.1, GEIH universe & 2001--2026: national survey (ECH and GEIH). 1993--2000: ICML reconstruction from annual survey levels of a different design. \\
Labor-informality rate & ICML 2.6.1, GEIH universe & 2021--2026: current definition. 2007--2020: inversion of published moving windows, with a bridge between definitions. 2001--2006: six aggregate anchors. 1993--2000: ICML backcast. \\
Relative labor earnings (informal / formal) & GEIH microdata, author's processing & 2007Q1--2026Q2, quarterly, by position and unit size; missing 2008Q2--Q4, 2011 and 2020Q2. Historical informality classification until 2021, current definition from 2022. \\
\bottomrule
\end{tabularx}
\end{table}

This provenance should be kept in mind when reading the results. Before 2001 the quarterly dynamics of employment and informality come from the ICML reconstruction, and before 2007 the informality rate rests on a small number of anchors. The estimation uses every quarter, which is what the reconstruction was built for, and Section~\ref{sec:identification} shows how much each period contributes to what we learn about the size of the informal sector.

\subsection{Relative labor earnings}
\label{sec:earnings-data}

\subsubsection{Construction}

The earnings series comes from the monthly GEIH microdata from January 2007 to August 2026, processed by the author. Each employed person is classified as formal or informal with the same rules that underlie the informality series, a size-and-position proxy until 2021 and the social-security definition from 2022 on, and persons whose status cannot be determined are left unclassified rather than imputed. For workers with positive labor income, excluding unpaid family workers, we compute monthly weighted sums of income and of persons with the survey's national expansion factor. The quarterly ratio is the weighted mean income of informal workers divided by that of formal workers. Income is DANE's derived labor income from the main and secondary jobs from mid-2008 on, and before that the sum of the main-job wage or profit and secondary-job income. An alternative ratio uses income per hour worked.

\subsubsection{Worker groups and value added}

The labor-only accounting of Section~\ref{sec:model} treats informal income as value added, so the series is also computed by group of worker. Over 2007--2026, 68.6\% of paid informal workers are own-account workers or employers on average, 22.3\% are employees in units of up to five persons, 5.6\% are domestic employees and 3.4\% are employees of units with six or more persons. All employees answer the size question (P6870 until 2021 and P3069 afterwards).

What the survey income represents differs across these groups. The income of own-account workers and employers is mixed income, whereas an employee's wage is labor compensation and leaves out the rest of the enterprise's value added. The baseline therefore uses a group-specific \emph{proxy} for informal value added per worker. For the first three groups it is reported income, and for employees of larger units it is the wage divided by $1-\alpha$, on the assumption that they work with the firm's capital. Treating the earnings of domestic workers and of employees of small units as all the value added attributable to them is an identifying assumption, since the survey does not observe it. The proxy is divided by the mean wage of formal employees, and formal own-account workers are left out of the denominator because their income is not a wage. The adjustment raises the sample-average ratio from 0.421 to 0.441, and Section~\ref{sec:limitations} returns to it.

\subsubsection{Non-wage labor costs}

The survey asks wage earners for their pay ``before deductions'', so the employee's own contributions are part of the reported wage. What separates the reported wage from the employer's total labor cost, which is what the model's gross formal wage $(1-\alpha)Y_f/N_f$ measures, are the employer's social-security contributions, the payroll contributions (\emph{parafiscales}) and the benefits paid outside the monthly wage. All of them are set by law, and we compute them quarter by quarter for a worker earning less than ten minimum wages. They are the employer's pension contribution (11.6\% in 2007 and 12\% from 2008), health (8.5\%), occupational risk (0.5\%), the family compensation fund (4\%), the payroll contributions to SENA and ICBF (5\%) and the benefits, that is, the service bonus, severance pay and its interest, and holidays (21.8\%). The 2012 tax reform (Law 1607) eliminated the SENA and ICBF contributions for these workers from May 2013 and the employer's health contribution from January 2014, so non-wage costs fall from 51.9\% of the wage in 2008--2012 to 38.4\% from 2014 on. Writing $d_t^{\mathrm{nw}}$ for the employer's non-wage cost premium over the reported wage, the log gross-up $\ln(1+d_t^{\mathrm{nw}})$ converts the reported formal wage into labor cost in the measurement equation of Section~\ref{sec:identification}. A sensitivity uses contributions only, which would be appropriate if reported income already included the benefits.

\subsubsection{The series}

\begin{figure}[H]
\centering
\includegraphics[width=0.85\textwidth]{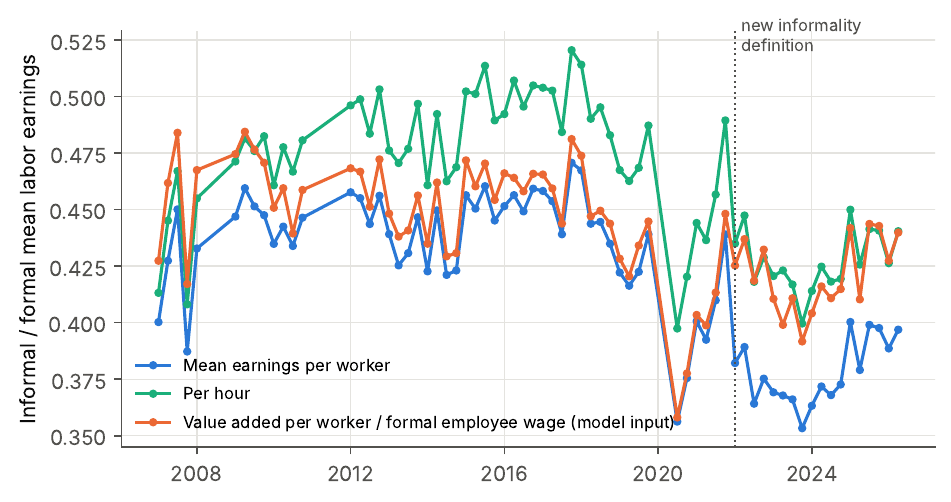}
\caption{Ratio of mean labor earnings of informal to formal workers, GEIH, 2007Q1--2026Q2. The dotted line marks the change in the definition of informality.}
\label{fig:earnings}
\end{figure}

The ratio moves between 0.42 and 0.47 from 2007 to 2019, falls during the pandemic and drops to about 0.37 when the current definition of informality comes into use in 2022 (Figure~\ref{fig:earnings}). The drop coincides with the change of definition, when the share of unclassified workers falls from about 3.5\% to almost zero, and the measurement equation treats it as a level shift. Twenty-one months of 2008 and 2011 have no usable expansion factor in the normalized files and are missing, and so are March and April 2020, when telephone interviewing left a quarter of the employed without a formality status. A quarter is computed only when its three months are valid.

\subsection{Transformations}

Let $Y_t$, $C_t$, $I_t$, $G_t$ and $N_t$ denote the levels and $s_t$ the informality rate. The observables are
\begin{equation}
\bm{y}^{\mathrm{obs}}_t = \big(\Delta\ln Y_t,\ \Delta\ln C_t,\ \Delta\ln I_t,\ \Delta\ln G_t,\ \Delta\ln N_t,\ \Delta\logit s_t\big)^{\top} - \bm{m}_{q(t)},
\label{eq:obs}
\end{equation}
where $\bm{m}_{q}$ is the mean of each growth rate in quarter of the year $q\in\{1,2,3,4\}$ over 2005--2018. Subtracting quarter-specific means removes deterministic seasonality and the mean growth rate at the same time, and the same means are used for the whole sample. The earnings ratio enters in logarithmic \emph{levels}, $\ln R_t$, because it is its level that identifies relative productivity. In the labor-only estimates the logit of the informality rate also enters in levels, measured from the logit of its sample mean after removing the quarter-of-year pattern of 2005--2018 (Section~\ref{sec:identification}). The sample runs from 1993Q2 to 2026Q2, 133 quarters in all. We treat 2020Q2--2020Q4 as missing in the likelihood, since those quarters combine an unprecedented fall in output and employment with a change in the way the survey was collected. This spares a first-order Gaussian approximation from having to fit such exceptional observations, at the cost of leaving the pandemic adjustment itself unmodeled.

Two features of the data shape the estimation. Quarterly growth rates have almost no persistence. Leaving out the pandemic, the first-order autocorrelation is $-0.13$ for GDP, $0.06$ for consumption and $-0.21$ for employment, so no model can forecast them much better than their mean. In addition, the growth rates of investment and of the logit informality rate are negatively autocorrelated ($-0.45$ and $-0.37$), which is what noise in levels produces.

\section{Model}
\label{sec:model}

The model is a standard real business-cycle economy with two ways of producing. A formal sector uses capital and registered labor, and an informal sector of own-account work and small units uses labor only. We keep it as close as possible to the textbook structure, so that the identification results cannot be attributed to ad hoc features. This section derives the equilibrium conditions from the households' and firms' problems, explains what each of them says and shows how the informal share of output follows from them. The stationary system that is solved numerically, the steady-state algorithm and the linear solution are collected in Appendix~\ref{app:model}.

\subsection{Overview and notation}

Time is quarterly. The economy has a representative household, competitive formal firms, informal production units owned and run by the household, and a government. Table~\ref{tab:notation} lists the main symbols. Capital letters denote levels that grow with the economy; the corresponding lowercase letters, used from Section~\ref{sec:steady} on, denote the same variables divided by the trend.

\begin{table}[H]
\centering
\caption{Main notation.}
\label{tab:notation}
\footnotesize
\begin{tabularx}{\textwidth}{l X}
\toprule
\textbf{Symbol} & \textbf{Meaning} \\
\midrule
$C_t, I_t, G_t$ & Private consumption, investment, government consumption \\
$K_t$ & Capital stock available in quarter $t$ (all capital is used by formal firms) \\
$N_{f,t}, N_{i,t}$ & Formal and informal work \\
$s_t = N_{i,t}/(N_{f,t}+N_{i,t})$ & Labor-informality rate \\
$Y_{f,t}, Y_{i,t}, Y_t$ & Formal, informal and total output, $Y_t = Y_{f,t}+Y_{i,t}$ \\
$\theta_t = Y_{i,t}/Y_t$ & Informal share of output; $\theta$ without subscript is its steady state \\
$w_{f,t}, r_t$ & Formal wage and rental rate of capital \\
$\tau_t$ & Formality wedge on formal labor income (taxes, contributions, compliance costs) \\
$X_t$ & Labor-augmenting technology, common to both sectors, with a unit root \\
$R_t$ & Informal income per worker relative to the gross formal wage \\
$\alpha$ & Capital share of formal output \\
$\eta$ & Elasticity of informal output to informal work ($0<\eta<1$) \\
$\varphi$ & Inverse Frisch elasticity of labor supply \\
$\mu_t, z^g_t, \nu_t, \psi_t, \omega_t, x_t, z^r_t$ & Shocks: trend growth, government, investment efficiency, labor supply, wedge, absorption, informal productivity \\
\bottomrule
\end{tabularx}
\end{table}

\subsection{Household}
\label{sec:household}

\subsubsection{Preferences}

A representative household chooses consumption, investment, next quarter's capital and the hours supplied to each sector to maximize expected lifetime utility
\begin{equation}
E_0 \sum_{t=0}^{\infty}\beta^t\Big[\ln C_t - \psi_t\,\frac{\chi_f N_{f,t}^{1+\varphi} + \chi_i N_{i,t}^{1+\varphi}}{1+\varphi}\Big].
\label{eq:utility}
\end{equation}
The first term is the utility of consumption, and the logarithm makes the value of an extra peso fall as consumption rises. The second term is the disutility of work. Each type of work has its own convex cost, so the hundredth hour of informal work is more costly than the first, and the same is true of formal work. This is the simplest specification in which the household supplies both types of work at once, as Colombian households do, rather than moving everyone to whichever sector pays slightly more. The weights $\chi_f$ and $\chi_i$ capture fixed differences in the attractiveness of the two sectors (access, flexibility, preferences) and are calibrated so that the model reproduces the mean informality rate. The shock $\psi_t$ shifts the disutility of all work (labor supply), $\beta$ is the discount factor and $1/\varphi$ is the Frisch elasticity of labor supply.

\subsubsection{Budget and capital accumulation}

Each quarter the household receives formal wages net of the wedge, the whole output of the informal units it runs, the rent on its capital, lump-sum transfers $T_t$ and an aggregate transfer $B_t$ for net external absorption, which it takes as given. It spends this income on consumption and investment,
\begin{equation}
C_t + I_t = (1-\tau_t)\,w_{f,t}N_{f,t} + Y_{i,t} + r_t K_t + T_t + B_t .
\label{eq:budget}
\end{equation}
Investment becomes capital with an efficiency that varies over time,
\begin{equation}
K_{t+1} = (1-\delta)K_t + \nu_t I_t ,
\label{eq:capital}
\end{equation}
where $\delta$ is depreciation and $\nu_t$ is a shock to the marginal efficiency of investment, which changes how much new capital a peso of investment buys, for instance because of financial conditions or the relative price of capital goods. The formality wedge is
\begin{equation}
1-\tau_t = (1-\bar\tau)\,e^{-\omega_t},
\label{eq:wedge}
\end{equation}
so that $\bar\tau$ is its steady-state level and an increase in $\omega_t$ makes formal work less attractive. The wedge stands for everything that separates what a formal job costs from what the worker takes home, including social-security contributions, payroll taxes and the costs of registration and compliance.

\subsubsection{First-order conditions}

Let $\lambda_t$ be the Lagrange multiplier of the budget constraint (the marginal utility of a peso) and $\lambda_tq_t$ that of the capital accumulation equation (the value of a unit of installed capital). The Lagrangian is
\begin{multline}
\mathcal{L} = E_0\sum_{t=0}^{\infty}\beta^t\Big\{\ln C_t - \psi_t\frac{\chi_fN_{f,t}^{1+\varphi}+\chi_iN_{i,t}^{1+\varphi}}{1+\varphi}
+ \lambda_t\big[(1-\tau_t)w_{f,t}N_{f,t} + Y_{i,t} + r_tK_t + T_t + B_t - C_t - I_t\big] \\
+ \lambda_tq_t\big[(1-\delta)K_t + \nu_tI_t - K_{t+1}\big]\Big\}.
\end{multline}
Setting the derivatives with respect to each choice to zero gives five conditions,
\begin{align}
C_t:&\quad \lambda_t = \frac{1}{C_t}, \label{eq:foc-c}\\
I_t:&\quad q_t = \frac{1}{\nu_t}, \label{eq:foc-i}\\
K_{t+1}:&\quad \lambda_tq_t = \beta E_t\big[\lambda_{t+1}\big(r_{t+1} + q_{t+1}(1-\delta)\big)\big], \label{eq:foc-k}\\
N_{f,t}:&\quad \psi_t\chi_fN_{f,t}^{\varphi} = \lambda_t(1-\tau_t)w_{f,t}, \label{eq:foc-nf}\\
N_{i,t}:&\quad \psi_t\chi_iN_{i,t}^{\varphi} = \lambda_t\,\frac{\partial Y_{i,t}}{\partial N_{i,t}} . \label{eq:foc-ni}
\end{align}
Condition \eqref{eq:foc-c} says that a peso is worth the extra utility of consuming it, and \eqref{eq:foc-i} that a unit of capital, which costs $1/\nu_t$ pesos of investment, has that price. Condition \eqref{eq:foc-k} says that the household gives up a unit of capital today, worth $\lambda_tq_t$, only if it expects to get the same value back tomorrow, discounted by $\beta$, from the rent $r_{t+1}$ and the undepreciated capital. Substituting \eqref{eq:foc-c} and \eqref{eq:foc-i} gives the Euler equation
\begin{equation}
\frac{1}{\nu_t C_t} = \beta E_t\Big[\frac{1}{C_{t+1}}\Big(r_{t+1} + \frac{1-\delta}{\nu_{t+1}}\Big)\Big].
\label{eq:euler}
\end{equation}
According to the two labor conditions, the household works in each sector until the utility cost of the last hour equals the utility value of what that hour produces. Replacing $\lambda_t$ with \eqref{eq:foc-c},
\begin{equation}
\psi_t\chi_f N_{f,t}^{\varphi} C_t = (1-\tau_t)\,w_{f,t}, \qquad \psi_t\chi_i N_{i,t}^{\varphi} C_t = \frac{\partial Y_{i,t}}{\partial N_{i,t}} .
\label{eq:foc-labor}
\end{equation}
The left-hand side is the marginal rate of substitution between leisure and consumption, that is, the consumption the household requires in exchange for one more hour of work. The right-hand side is what the hour pays, the after-wedge formal wage in one case and the marginal product of informal work in the other. A higher wedge lowers the right-hand side of the formal condition, and the household shifts hours towards informal work until both conditions hold again.

\subsection{Formal firms}

\subsubsection{Technology and factor prices}

Formal firms are competitive, rent capital from the household and hire formal work to produce with a Cobb--Douglas technology,
\begin{equation}
Y_{f,t} = K_t^{\alpha}\big(X_tN_{f,t}\big)^{1-\alpha}.
\label{eq:technology}
\end{equation}
A firm maximizes profits $Y_{f,t} - w_{f,t}N_{f,t} - r_tK_t$, which gives
\begin{equation}
r_t = \alpha\,\frac{Y_{f,t}}{K_t}, \qquad w_{f,t} = (1-\alpha)\,\frac{Y_{f,t}}{N_{f,t}} .
\label{eq:foc-firm}
\end{equation}
Each factor is paid its marginal product, and with constant returns these payments exhaust output, a share $\alpha$ going to capital and $1-\alpha$ to formal labor. We use this property to estimate $\alpha$ (Section~\ref{sec:alpha}), and it is also why the formal wage measured in the survey is a fraction $1-\alpha$ of formal value added per worker.

\subsubsection{Productivity trend}

Labor-augmenting technology has a unit root with stochastic growth \citep{AguiarGopinath2007},
\begin{equation}
\ln X_t - \ln X_{t-1} = \gamma + \mu_t ,
\label{eq:trend}
\end{equation}
so a shock $\mu_t$ shifts the level of productivity permanently. This is the usual way of representing the persistent swings in growth of emerging economies, and it gives output, consumption, investment and capital a common stochastic trend.

\subsection{Informal production}

\subsubsection{Technology}

Following the shadow-economy literature \citep{IhrigMoe2004, ElginOztunali2012}, informal production uses labor only, with decreasing returns,
\begin{equation}
Y_{i,t} = A_i\,e^{z^r_t}\,X_t\,N_{i,t}^{\eta}, \qquad 0<\eta<1 .
\label{eq:informal-technology}
\end{equation}
where $A_i$ is the level of informal productivity and $z^r_t$ a shock to it, and $X_t$ makes informal output grow with the economy. With decreasing returns ($\eta<1$) each additional informal worker adds less than the previous one, since informal activities compete for the same customers and spaces and cannot be scaled up without capital.

\subsubsection{Average and marginal products}

Because the household runs the informal unit itself, there is no separate informal wage, and two different quantities have to be kept apart,
\begin{equation}
\underbrace{\frac{Y_{i,t}}{N_{i,t}}}_{\text{average product: income per informal worker}}
\qquad\text{and}\qquad
\underbrace{\frac{\partial Y_{i,t}}{\partial N_{i,t}} = \eta\,\frac{Y_{i,t}}{N_{i,t}}}_{\text{marginal product: return of one more hour}} .
\label{eq:avg-marginal}
\end{equation}
The marginal product governs the household's choice of hours in \eqref{eq:foc-labor}. The average product is the latent value added per worker that we try to approximate with survey earnings, which need not coincide with the wage or profit reported by each respondent.

\subsubsection{Why informal production uses labor only}

Informal units do use capital. The reason for the labor-only technology lies in what the household survey measures. The income that own-account workers and informal employers report is mixed income or profit, which already includes the return to the vehicles, tools and premises of their unit, while the wages they pay show up in the income of their employees. Informal survey income is therefore a measure of informal value added per worker. The income of formal wage employees, by contrast, is only the labor share of formal value added, as \eqref{eq:foc-firm} shows. The labor-only technology carries these measurement conventions \emph{into the model}, in which informal income per worker is informal output per worker and the formal wage is a fraction $1-\alpha$ of formal output per worker. Employees of informal units are the one group whose wage does not include the unit's surplus, and the earnings series adjusts for them by group (Section~\ref{sec:earnings-data}). Other treatments appear in the sensitivity analysis.

\subsection{Government, resources and shocks}

\subsubsection{Government and the resource constraint}

Government consumption is exogenous and grows with the trend, $G_t = \bar g\,X_t\,e^{z^g_t}$, and the government balances its budget with lump-sum transfers, $T_t = \tau_tw_{f,t}N_{f,t} - G_t$. The wedge is thus a transfer between agents rather than a loss of resources, and its effect is on how work is allocated. To account for the national-account components that a closed three-good model leaves out, let $B_t=(\bar x+x_t)Y_t$ be a reduced-form net transfer that finances domestic absorption in excess of output (or, when negative, resources supplied to the rest of the world), which individual households take as given. It is an accounting closure, not a model of the external sector. Adding up the household and government budgets and using \eqref{eq:foc-firm} gives
\begin{equation}
Y_t\,\big(1 + \bar x + x_t\big) = C_t + I_t + G_t .
\label{eq:resources}
\end{equation}
The term $\bar x+x_t$ captures the omitted absorption and the discrepancy that arises because chained volumes are not additive; $Y=C+I+G$ holds neither in Colombia's open economy nor for chained-volume components. Its steady-state value $\bar x$ reproduces the sample expenditure ratios, and $x_t$ absorbs fluctuations in the omitted components. Since $B_t$ is never repaid, the closure says nothing about trade, external debt or fiscal financing.

\subsubsection{Shocks}

The seven exogenous processes are independent first-order autoregressions,
\begin{equation}
\zeta_t = \rho_\zeta\,\zeta_{t-1} + \sigma_\zeta\,\varepsilon_{\zeta,t},\qquad \varepsilon_{\zeta,t}\sim N(0,1),\qquad \zeta\in\{\mu,\, z^g,\, \ln\nu,\, \ln\psi,\, \omega,\, x,\, z^r\},
\label{eq:shocks}
\end{equation}
with persistence $\rho_\zeta$ and volatility $\sigma_\zeta$. The shock $\mu$ moves productivity growth, $z^g$ government spending, $\nu$ the efficiency of investment, $\psi$ the willingness to work, $\omega$ the cost of formality, $x$ absorption and $z^r$ informal productivity. The informal-productivity shock is active only when the earnings observable of Section~\ref{sec:identification} is used. Without that observable nothing in the data distinguishes it from the wedge, so it is switched off in the estimates that use aggregates only.

\subsubsection{Equilibrium}

An equilibrium is a set of quantities and prices that satisfy the household conditions \eqref{eq:capital}, \eqref{eq:euler} and \eqref{eq:foc-labor}, the firm conditions \eqref{eq:foc-firm}, the technologies \eqref{eq:technology} and \eqref{eq:informal-technology}, the resource constraint \eqref{eq:resources} and the laws of motion of the shocks. Once prices are substituted out, the model has six endogenous variables ($C$, $Y$, $I$, $N_f$, $N_i$, $K$) and seven shocks, for thirteen equations in all (Appendix~\ref{app:model}).

\subsection{The informal share of output}

\begin{proposition}[Informal share of output]
\label{prop:share}
Let $R_t$ be informal income per worker relative to the gross formal wage, $R_t \equiv (Y_{i,t}/N_{i,t})/w_{f,t}$, and $s_t \equiv N_{i,t}/(N_{f,t}+N_{i,t})$ the informality rate. Then
\begin{equation}
\theta_t \equiv \frac{Y_{i,t}}{Y_t} = \frac{(1-\alpha)\,R_t\,o_t}{1+(1-\alpha)\,R_t\,o_t},\qquad o_t\equiv\frac{s_t}{1-s_t}.
\label{eq:share}
\end{equation}
If instead both sectors shared the Cobb--Douglas technology with mobile capital, the factor $1-\alpha$ would disappear and $\theta_t$ would equal the informal share of labor income, $R_to_t/(1+R_to_t)$.
\end{proposition}

\emph{Proof.} Write the ratio of informal to formal output as the product of relative output per worker and relative employment,
\[
\frac{Y_{i,t}}{Y_{f,t}} = \frac{Y_{i,t}/N_{i,t}}{Y_{f,t}/N_{f,t}}\cdot\frac{N_{i,t}}{N_{f,t}}
= \frac{R_t\,w_{f,t}}{w_{f,t}/(1-\alpha)}\cdot o_t = (1-\alpha)\,R_t\,o_t ,
\]
where $o_t$ is the odds of informal employment and the second step uses the definition of $R_t$, the wage $w_{f,t}=(1-\alpha)Y_{f,t}/N_{f,t}$ from \eqref{eq:foc-firm} and $N_{i,t}/N_{f,t}=s_t/(1-s_t)$. Then $\theta_t = (Y_{i,t}/Y_{f,t})/(1+Y_{i,t}/Y_{f,t})$. With a common technology both sectors have the same capital per efficiency unit of labor, so output is proportional to labor income in each of them and the factor $1-\alpha$ cancels. $\square$

Proposition~\ref{prop:share} is an accounting identity of the model that splits the measurement problem into three parts. The informality rate $s_t$ is observed directly in the household survey. The earnings ratio $R_t$ is the quantity that has to be identified, and it too can be measured in the survey, as mean informal income over the mean wage of formal employees once the wedge is removed. The capital share $\alpha$, which converts the formal wage into formal output per worker, is estimated from the national accounts.

A numerical example shows how much each part matters. With an informality rate of 55\% ($o=1.22$) and $\alpha=0.35$, an earnings ratio of $R=0.2$ gives $\theta=14\%$ and $R=0.6$ gives $\theta=32\%$. Holding $R$ at 0.35, raising $\alpha$ from 0.35 to 0.50 lowers $\theta$ from 22\% to 18\%, whereas assuming equal technologies raises it to 30\%. The equal-technology expression omits the factor $1-\alpha$ and is therefore larger for any given $R$ and $s$, which makes it an upper bound for given data.

\subsection{What moves the informality rate}

Dividing the two labor conditions in \eqref{eq:foc-labor} eliminates consumption and the labor-supply shock, which affect both sectors alike, and leaves the relative return of the two types of work,
\begin{equation}
\frac{\chi_i}{\chi_f}\Big(\frac{N_{i,t}}{N_{f,t}}\Big)^{\varphi} = \frac{\eta\,Y_{i,t}/N_{i,t}}{(1-\tau_t)(1-\alpha)\,Y_{f,t}/N_{f,t}} .
\label{eq:relative-return}
\end{equation}
The household moves work towards the informal sector when the informal marginal return rises relative to the after-wedge formal wage, and the convexity of the cost of work ($\varphi>0$) limits how far it goes.

\begin{proposition}[Determinants of the informality rate]
\label{prop:rate}
Taking logarithms of \eqref{eq:relative-return}, with $Y_{i,t}/N_{i,t} = A_ie^{z^r_t}X_tN_{i,t}^{\eta-1}$,
\begin{equation}
\varphi\logit s_t + (1-\eta)\ln N_{i,t} + \ln\frac{Y_{f,t}}{X_tN_{f,t}}
= \ln\frac{\chi_f\,\eta A_i}{\chi_i\,(1-\alpha)} - \ln(1-\bar\tau) + z^r_t + \omega_t .
\label{eq:rate-labor-only}
\end{equation}
The right-hand side contains only the wedge and informal productivity. The left-hand side contains, besides the informality rate, the scale of informal work and formal output per efficiency unit of labor, which depends on capital per formal worker. Aggregate shocks to productivity, investment efficiency, labor supply, government spending or absorption therefore move the informality rate only through capital per formal worker and through the scale of informal employment. With equal technologies those two terms disappear and
\begin{equation}
\varphi\,\logit s_t = \ln\frac{\chi_f}{\chi_i} - \ln(1-\bar\tau) + \ln\tilde\rho + z^r_t + \omega_t ,
\label{eq:rate}
\end{equation}
so aggregate shocks do not move the informality rate at all.
\end{proposition}

\emph{Proof.} Take logarithms of \eqref{eq:relative-return}, use $N_{i,t}/N_{f,t} = s_t/(1-s_t)$, $\ln[s_t/(1-s_t)]=\logit s_t$ and \eqref{eq:wedge}, and collect the endogenous terms on the left. With equal technologies $Y_t=K_t^{\alpha}(X_tL_t)^{1-\alpha}$, $L_t = N_{f,t}+\tilde\rho e^{z^r_t}N_{i,t}$, the informal wage is $\tilde\rho e^{z^r_t}w_{f,t}$ and the ratio of wages does not depend on capital or on the scale of either sector. $\square$

One empirical consequence of Proposition~\ref{prop:rate} is that the estimated variance of the informality rate is attributed mostly to the formality wedge and to informal productivity (Section~\ref{sec:results}). It is also why the informality rate carries little information about $R_t$ beyond its own level.

\subsection{An alternative: equal technologies}
\label{sec:equal-tech}

For comparison, and to obtain an upper bound, we also estimate the model with a common Cobb--Douglas technology, $Y_t=K_t^{\alpha}(X_tL_t)^{1-\alpha}$ with $L_t=N_{f,t}+\tilde\rho e^{z^r_t}N_{i,t}$, mobile capital and an informal wage $\tilde\rho e^{z^r_t}w_{f,t}$, where $\tilde\rho$ is the relative productivity of informal work. In that case $R_t = \tilde\rho e^{z^r_t}$ and the share of output equals the share of labor income.

\subsection{Trend, steady state and solution}
\label{sec:steady}

\subsubsection{Detrending}

Output, consumption, investment, government consumption and capital grow with $X_t$ and have no constant steady state. We make the model stationary by dividing them by the trend, with $y_t=Y_t/X_t$, $c_t=C_t/X_t$ and so on, and $k_t=K_t/X_{t-1}$ for capital, which is chosen a quarter in advance. Work and the informality rate are stationary already. Appendix~\ref{app:model} lists the stationary equations.

\subsubsection{Steady state}

With all shocks at zero, the stationary model has a steady state that can be computed in closed form for a given $\theta$. The Euler equation \eqref{eq:euler} with constant consumption growth $e^{\gamma}$ gives the rental rate, $r = e^{\gamma}/\beta - 1 + \delta$, which is 3.5\% a quarter under the calibration of Section~\ref{sec:estimation}. The firm condition \eqref{eq:foc-firm} then gives the capital--output ratio of the formal sector, $K/Y_f = \alpha/r$, equal to 10.0 quarters or 2.5 years, close to the 2.4--2.8 years that DANE measures for the whole economy. Hours are split according to the mean informality rate, $N_f = \bar N(1-\bar s)$ and $N_i=\bar N\bar s$, where $\bar N$ is a normalization. Formal output follows from the technology, $Y_f = (K/Y_f)^{\alpha/(1-\alpha)}N_f$ in detrended units, and total output from the informal share, $Y = Y_f/(1-\theta)$, so that $Y_i=\theta Y$ and $A_i = Y_i/N_i^{\eta}$. Investment replaces depreciation and keeps capital growing with the trend, $I = (e^{\gamma}-1+\delta)\,\tilde K$ with $\tilde K = K_t/X_t$. The implied ratio $I/Y$ of 14\% is below the 19\% in the data because the model has no residential or public investment; consumption and government consumption are set at their sample ratios to GDP, and $\bar x$ closes the resource constraint. Finally, $\chi_f$ and $\chi_i$ take whatever values make the labor conditions \eqref{eq:foc-labor} hold at the chosen $N_f$ and $N_i$.

This last step matters for identification. Since $A_i$, $\chi_f$ and $\chi_i$ adjust, \emph{any} value of $\theta$ is consistent with the observed informality rate and with every steady-state ratio, and the first moments of the aggregates do not pin it down. The same is true with equal technologies, where $\tilde\rho=\theta(1-\bar s)/[(1-\theta)\bar s]$. Whatever the data say about $\theta$ has to come from the dynamics or from an additional observable (Section~\ref{sec:identification}).

\subsubsection{Solution}

We linearize the thirteen stationary equations around the steady state, replacing each variable by its log deviation and each equation by its first-order Taylor approximation. The result is a linear system $\bm{A}\,E_t\bm{z}_{t+1} = \bm{B}\,\bm{z}_t$ in which eight variables are predetermined (capital and the seven exogenous processes) and the other five (consumption, output, investment and the two types of work) can jump when news arrives. The generalized Schur decomposition \citep{Klein2000} yields the unique stable solution
\begin{equation}
\bm{s}_{t+1} = \bm{T}\bm{s}_t + \bm{R}\bm{\varepsilon}_{t+1},\qquad \text{jumps}_t = \bm{P}\bm{s}_t ,
\label{eq:solution}
\end{equation}
where $\bm{s}_t$ collects the eight state variables. The solution exists and is unique when the number of stable eigenvalues equals the number of states (the Blanchard--Kahn condition). We check this at every parameter draw and give zero posterior density to draws that fail it.

\section{Estimation}
\label{sec:estimation}

\subsection{State-space form}

Let $\bm{x}_t$ be the vector of predetermined variables of the linearized model (capital and the exogenous processes), with law of motion $\bm{x}_{t+1} = \bm{T}\bm{x}_t + \bm{R}\bm{\varepsilon}_{t+1}$, and let all model variables be $\bm{M}\bm{x}_t$. Each observable in \eqref{eq:obs} is a difference of model variables plus the trend shock for the variables with a trend, for example $\Delta\ln Y_t = \hat y_t - \hat y_{t-1} + \mu_t$. The state is therefore augmented with its lag.

Survey and statistical noise is modeled in \emph{levels}. For investment, employment and the logit informality rate, the observed log level is the model level plus an independent error $e_{j,t}\sim N(0,\sigma^2_{e_j})$, so the growth rate contains $e_{j,t}-e_{j,t-1}$, and these errors are carried as states. The specification reproduces the negative autocorrelation of the corresponding growth rates and lets the Kalman smoother separate survey noise from movements in the latent informality rate. We also add a small error, with standard deviation $10^{-3}$, to every growth rate to keep the innovation covariance non-singular. The likelihood is computed with the Kalman filter, initialized at the unconditional distribution, and quarters with missing observations skip the update step \citep{DurbinKoopman2012}. We checked the implementation against the exact Gaussian likelihood of the stacked observations. Appendix~\ref{app:model} writes out every measurement equation and describes the filter and the sampler in detail.

\subsection{Calibration}

Table~\ref{tab:calibration} lists the parameters fixed before estimation. Those taken from the data are sample means for 1993--2026, and the capital share and depreciation come from Colombian sources and from the data, as described below. The equal-technology estimates use $\alpha=0.33$, for which the depreciation implied by the investment ratio is also 4\% a year. The informal share $\theta$ is estimated rather than calibrated, and the steady state is recomputed at every draw; Section~\ref{sec:identification} shows that the path of the informal share, which is what we are after, does not depend on its value.

\begin{table}[H]
\centering
\caption{Calibrated parameters.}
\label{tab:calibration}
\footnotesize
\begin{tabularx}{\textwidth}{L{1.6cm} L{6.2cm} r Y}
\toprule
\textbf{Parameter} & \textbf{Meaning} & \textbf{Value} & \textbf{Source} \\
\midrule
$\alpha$ & Capital share of formal output & 0.354 & Estimated, Section~\ref{sec:alpha} \\
$\beta$ & Quarterly discount factor & 0.985 & Conventional, emerging economy \\
$\varphi$ & Inverse Frisch elasticity & 1.0 & Conventional \\
$\bar\tau$ & Steady-state formality wedge & 0.25 & Affects only the scale of $\chi_f$ \\
$\gamma$ & Mean quarterly trend growth & 0.0096 & GDP, 2005--2018 \\
$\bar s$ & Mean informality rate & 0.645 & ICML, 1993--2026 \\
$C/Y$, $I/Y$, $G/Y$ & Expenditure ratios & 0.697, 0.186, 0.139 & National accounts, 1993--2026 \\
$\delta$ & Quarterly depreciation & 0.0102 & 4\% a year; GRECO 3.1\%, DNP 3.8\% \\
$\bar x$ & Chain-volume residual & -- & Implied by expenditure ratios \\
$\ln(1+d_t^{\mathrm{nw}})$ & Non-wage labor costs of formal employees & 0.42 / 0.32 & Statutory rates, 2008--2012 / 2014--; Section~\ref{sec:data} \\
\bottomrule
\end{tabularx}
\end{table}

\subsection{Capital share and depreciation}
\label{sec:alpha}

\subsubsection{Depreciation}

Depreciation is set at the standard 4\% a year, which is close to the Colombian estimates. The capital stocks of the Banco de la Rep\'ublica growth project (GRECO), built with the method of \citet{Barrios1993}, imply 2\% for buildings, 5\% for machinery and equipment and 3.1\% for total fixed capital between 1950 and 1997 \citep{GRECO1999}, and the National Planning Department uses 3.8\% \citep{RodriguezPerillaReyes2004}.

\subsubsection{Capital share}

The capital share of formal output is estimated. We build capital by perpetual inventory from gross fixed capital formation at constant prices and estimate
\begin{equation}
\ln\frac{Y_{f,t}}{N_{f,t}} = c + \alpha\ln\frac{K_t}{N_{f,t}} + g\,t + e_t
\label{eq:dols}
\end{equation}
by dynamic OLS \citep{StockWatson1993}, with one lead and one lag and Newey--West errors, on annual data for 1993--2025. Because formal output $Y_{f,t}=(1-\theta_t)Y_t$ and formal employment $N_{f,t}=(1-s_t)N_t$ depend on $\alpha$ through \eqref{eq:share}, the estimate is a fixed point. With 4\% depreciation it is 0.354 (standard error 0.25), and it falls to 0.33 with 6\% and 0.30 with 8\%. Without the trend the regression gives 0.39--0.43, because capital then absorbs productivity growth, and the aggregate regression gives an implausible 0.65. The series is short and the estimate imprecise, but it lies in the range of earlier Colombian estimates. GRECO finds 0.42 for 1925--1981 by cointegration and 0.36--0.41 for 1926--1994 and 1950--1994 by restricted least squares \citep{GRECO1999}, and the National Planning Department finds 0.39--0.51 depending on the specification \citep{RodriguezPerillaReyes2004}. Section~\ref{sec:results} reports sensitivities with $\alpha=0.33$ (with 6\% depreciation) and $\alpha=0.50$.

\subsection{Priors and sampler}

\subsubsection{Priors}

The estimated parameters are the informal share $\theta$, the labor elasticity of informal production $\eta$ (labor-only model), the seven persistences and seven innovation standard deviations (including the relative-productivity shock), the four measurement-error standard deviations (three level errors and the error of the earnings ratio) and the level shift $b$ of the earnings ratio from 2022Q1. Persistences, $\theta$ and $\eta$ have beta priors ($\eta$ with mean 0.7 and standard deviation 0.1), standard deviations inverse-gamma priors and $b$ a normal prior with mean zero and standard deviation 0.2 (Table~\ref{tab:posterior}). The prior of $\theta$, with mean 0.35 and standard deviation 0.10, is deliberately wide, and its 90\% interval covers informal shares from about 19\% to 52\% of output.

\subsubsection{Sampler and convergence}

We sample the posterior in an unconstrained parameterization (logit for parameters in the unit interval, logarithm for standard deviations and identity for $b$), including the Jacobian. The mode is found from three starting points by Nelder--Mead followed by BFGS, and the inverse Hessian at the mode, computed by finite differences, serves as the proposal covariance of a random-walk Metropolis--Hastings sampler \citep{AnSchorfheide2007}. Eight chains start from over-dispersed points around the mode. During 15{,}000 burn-in iterations the proposal scale is adapted towards an acceptance rate of 0.28 \citep{RobertsGelmanGilks1997}, after which 40{,}000 iterations per chain are kept with a fixed proposal. The equal-technology and aggregates-only estimates use four chains of 30{,}000 draws after 10{,}000. Convergence is assessed with the rank-normalized split $\hat R$ and the bulk and tail effective sample sizes of \citet{Vehtari2021}.

\section{Identification of the size of the informal sector}

This section first asks what aggregate data alone can say about the level of the informal share, using the model without the earnings observable, and then introduces the observable that identifies it. We report the evidence for the equal-technology model, where the mechanism is easiest to see; the labor-only technology leads to the same conclusion (Section~\ref{sec:alt-tech}).
\label{sec:identification}

\subsection{Where information about the level can come from}

Section~\ref{sec:model} shows that steady-state ratios are consistent with any value of $\theta$. In the log-linearized equal-technology model, $\theta$ enters the dynamics only through the weights of the two types of work in effective labor,
\begin{equation}
d\ln L_t = (1-\theta)\,d\ln N_{f,t} + \theta\,d\ln N_{i,t},
\label{eq:effective}
\end{equation}
so the only source of information is how output co-moves with the composition of employment. When the wedge moves workers between sectors, output changes more if informal work is much less productive than formal work (a small $\theta$) and less if the two are similar (a large $\theta$). This is a single and weak moment, and it is measured with noise because the informality rate itself is noisy (Section~\ref{sec:data}).

\subsection{Evidence from aggregate data}

We apply three standard checks, namely the profile likelihood, the dispersion of the posterior relative to the prior and the recovery of a known value in simulated data.

\subsubsection{Profile likelihood}

\begin{figure}[H]
\centering
\includegraphics[width=0.85\textwidth]{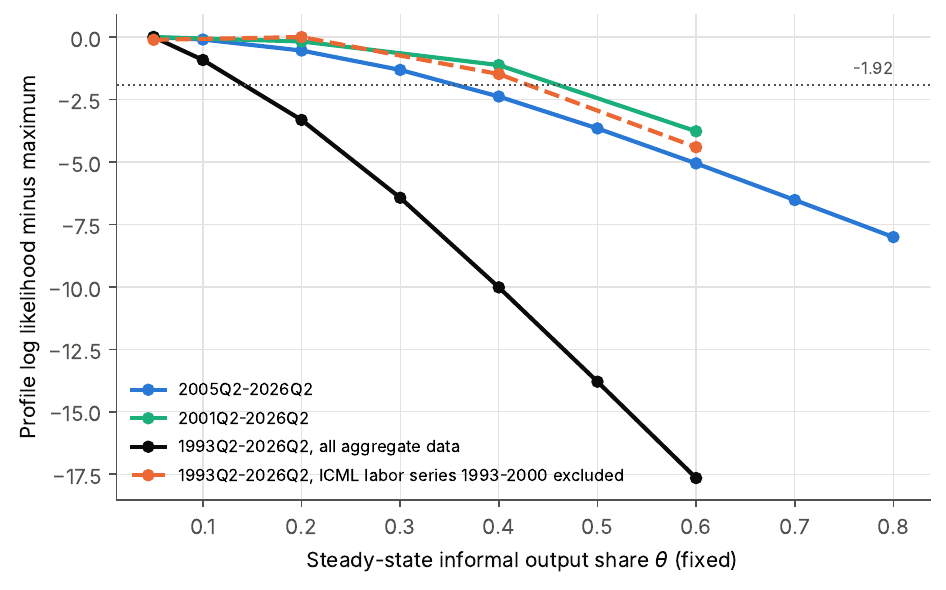}
\caption{Profile log likelihood of the steady-state informal share $\theta$, by sample. Values relative to the maximum of each profile; the dotted line marks $-1.92$.}
\label{fig:profiles}
\end{figure}

Figure~\ref{fig:profiles} and Table~\ref{tab:profiles} show the profile log likelihood, obtained by fixing $\theta$ at each value of a grid, re-optimizing the other parameters and reporting the maximized log likelihood relative to its overall maximum. A drop of 1.92 is the usual reference, corresponding to a one-parameter 95\% likelihood-ratio interval under regular identification. Since weak identification is exactly what is in question here, we use the line as a descriptive benchmark and not as a formal test.

\begin{table}[H]
\centering
\caption{Profile log likelihood relative to its maximum.}
\label{tab:profiles}
\footnotesize
\begin{tabularx}{\textwidth}{Y rrrr}
\toprule
\textbf{Sample} & $\theta=0.05$ & $\theta=0.20$ & $\theta=0.40$ & $\theta=0.60$ \\
\midrule
1993Q2--2026Q2, all aggregate data & 0.00 & $-3.32$ & $-10.02$ & $-17.65$ \\
1993Q2--2026Q2, ICML informality 1993--2000 left out & 0.00 & $-3.93$ & $-13.82$ & $-18.76$ \\
1993Q2--2026Q2, ICML employment and informality 1993--2000 left out & $-0.11$ & 0.00 & $-1.48$ & $-4.41$ \\
2001Q2--2026Q2 & 0.00 & $-0.18$ & $-1.12$ & $-3.77$ \\
2005Q2--2026Q2 & 0.00 & $-0.54$ & $-2.38$ & $-5.06$ \\
\bottomrule
\end{tabularx}
\end{table}

With the full sample the likelihood looks informative, favoring small informal shares and rejecting values above about 0.15. That information comes from 1993--2000, and more precisely from the employment series of those years. Leaving out the informality rate for 1993--2000 does not change the profile, but leaving out employment as well makes it nearly flat, like the profiles of the samples that start in 2001 or 2005. In those samples the likelihood varies by less than 1.5 log points between $\theta=0.05$ and $\theta=0.40$ and only begins to penalize the largest values.

\subsubsection{Prior and posterior dispersion}

With all the aggregate data, the posterior standard deviation of $\theta$ is 57\% of the prior one (median 0.18, 90\% interval 0.10--0.29). It is 80\% when the 1993--2000 labor series are left out and 81\% in the sample that starts in 2005.

\subsubsection{Simulated data}

We simulate the model at the posterior median of the 2005 sample, with a true informal share of 0.45, ten points above the prior mean, over the same 85 quarters and with the same missing quarters, and re-estimate it (posterior mode and Laplace standard deviation) in three replications. Table~\ref{tab:simulation} reports the results. With the six aggregates the estimate lies between 0.51 and 0.56 with a standard deviation of about 0.08, close to the prior's 0.10, so the data do not recover the true value.

\begin{table}[H]
\centering
\caption{Recovery of a known informal share ($\theta=0.45$) in simulated data, 2005Q2--2026Q2.}
\label{tab:simulation}
\footnotesize
\begin{tabularx}{\textwidth}{Y rrr rrr}
\toprule
& \multicolumn{3}{c}{\textbf{Six aggregates}} & \multicolumn{3}{c}{\textbf{Plus relative earnings}} \\
\cmidrule(lr){2-4}\cmidrule(lr){5-7}
\textbf{Replication} & 1 & 2 & 3 & 1 & 2 & 3 \\
\midrule
Posterior mode of $\theta$ & 0.561 & 0.508 & 0.554 & 0.450 & 0.448 & 0.459 \\
Laplace standard deviation & 0.083 & 0.076 & 0.079 & 0.004 & 0.004 & 0.005 \\
Ratio to prior standard deviation & 0.83 & 0.76 & 0.79 & 0.04 & 0.04 & 0.05 \\
\bottomrule
\end{tabularx}
\end{table}

\subsection{Alternative technology}
\label{sec:alt-tech}

A common alternative in the shadow-economy literature gives the informal sector a labor-only technology with decreasing returns, $Y_{i,t} = A_i X_t N_{i,t}^{\eta}$ \citep{IhrigMoe2004, ElginOztunali2012}. Aggregate shocks then change the relative return of the two types of work, and the informality rate responds to capital and to the trend shock. Estimated on the 2005 sample with $\eta$ free (prior mean 0.7), however, the profile likelihood of $\theta$ is as flat as with the common technology, falling by only 2.1 log points between $\theta=0.10$ and $\theta=0.40$. Changing the technology does not solve the identification problem.

\subsection{An observable that identifies the level}

\subsubsection{The earnings measurement equation}

Proposition~\ref{prop:share} points to what is missing, which is a direct measure of the earnings ratio $R_t$. In the labor-only model the observed ratio of informal income per worker to the mean wage of formal employees is
\begin{equation}
\ln R^{\mathrm{obs}}_t = \ln\frac{Y_{i,t}}{N_{i,t}} - \ln\Big((1-\alpha)\frac{Y_{f,t}}{N_{f,t}}\Big) + \ln(1+d_t^{\mathrm{nw}}) + b\,\mathbb{1}_{t\ge 2022Q1} + e^{R}_t,
\label{eq:earnings}
\end{equation}
where $\ln(1+d_t^{\mathrm{nw}})$ converts the wage reported by formal employees into the employer's labor cost using the statutory non-wage costs of Section~\ref{sec:data} (a constant wedge, $-\ln(1-\bar\tau)$ with $\bar\tau=0.25$, in a sensitivity), $b$ is a level shift for the change in the definition of informality and $e^{R}_t$ is measurement error. In log deviations the first two terms are $z^r_t + (\eta-1)\hat n_{i,t} - \alpha(\hat k_t - \mu_t - \hat n_{f,t})$, and their steady-state value is a function of $\theta$. With equal technologies the first two terms reduce to $\ln\tilde\rho + z^r_t$. The ratio can be computed from the household survey for every month since 2007 with the same classification of informality as the employment series. Equation \eqref{eq:earnings} enters the state space in levels and activates the shock $z^r_t$. In the simulation of Table~\ref{tab:simulation}, adding this observable recovers the true share with a standard deviation of half a percentage point.

\subsubsection{The informality rate in levels}

In the labor-only model the earnings ratio depends on informal employment through $\eta$. If the informality rate entered only in changes, a very persistent deviation of the wedge could shift $\hat n_{i,t}$ and mimic a different $\theta$, and in a first estimation the chains did settle on different values of $\theta$ ($\hat R=1.57$). The labor-only estimates therefore observe the logit of the informality rate in levels, measured from the logit of its sample mean after removing the quarter-of-year pattern and with an independent level error. This pins down the state of the wedge in every quarter and removes the trade-off.

\subsection{Identification in the baseline model}
\label{sec:id-baseline}

We now repeat the checks for the baseline, the labor-only model with the informality rate in levels and the earnings observable.

\subsubsection{Recovery of a known share}

We simulate data from the baseline at its posterior median with a known steady-state share of 0.20 or 0.40, with the same length and missing quarters as the actual data, and re-estimate $\theta$ (posterior mode and Laplace standard deviation) with and without the earnings observable (Table~\ref{tab:sim-labor-only}). Without earnings the estimate is roughly centered on the true value, but its standard deviation is 41\%--46\% of the prior's. With earnings the share is recovered within half a percentage point, with a standard deviation of 3\%--5\% of the prior's.

\begin{table}[H]
\centering
\caption{Recovery of a known steady-state share in data simulated from the baseline model (three replications per value).}
\label{tab:sim-labor-only}
\footnotesize
\begin{tabular}{c l c c c c}
\toprule
\textbf{True $\theta$} & \textbf{Observables} & \textbf{Mean mode} & \textbf{sd of modes} & \textbf{Laplace sd} & \textbf{Ratio to prior sd} \\
\midrule
0.20 & Aggregates only & 0.184 & 0.012 & 0.046 & 0.46 \\
0.20 & With earnings ratio & 0.203 & 0.003 & 0.003 & 0.03 \\
0.40 & Aggregates only & 0.407 & 0.040 & 0.041 & 0.41 \\
0.40 & With earnings ratio & 0.402 & 0.005 & 0.005 & 0.05 \\

\bottomrule
\end{tabular}
\end{table}

\subsubsection{Profile likelihood in the data}

The actual data behave differently (Table~\ref{tab:profile-labor-only}). With or without earnings, the profile likelihood of the steady-state share peaks at the lowest value of the grid and falls monotonically. The likelihood alone does not pin down the steady state, because the data can be matched with a low $\theta$ and an economy that stays away from its steady state for the whole sample, through unobserved levels such as the capital stock and total employment relative to their long-run values, which very persistent shocks can sustain. Observing the capital--output ratio built by perpetual inventory does not change this (Section~\ref{sec:results}).

\begin{table}[H]
\centering
\caption{Profile log likelihood of the steady-state share, baseline model with the constant wedge (relative to the maximum).}
\label{tab:profile-labor-only}
\footnotesize
\begin{tabular}{c c c}
\toprule
$\theta$ & \textbf{Aggregates only} & \textbf{With earnings ratio} \\
\midrule
0.100 & 0.00 & 0.00 \\
0.150 & -1.94 & -2.35 \\
0.200 & -4.11 & -4.68 \\
0.250 & -6.75 & -7.95 \\
0.277 & -8.36 & -10.22 \\
0.300 & -9.83 & -11.92 \\
0.350 & -13.51 & -17.86 \\
0.400 & -17.84 & -25.15 \\
0.450 & -22.97 & -33.06 \\

\bottomrule
\end{tabular}
\end{table}

\subsubsection{What the data identify}

What we want to measure, however, is the informal share in each quarter rather than its steady state. By Proposition~\ref{prop:share}, that share depends on the informality rate and the earnings ratio of the quarter, both of which are measured, and on $\alpha$. Table~\ref{tab:invariance} fixes the steady-state share at values between 0.10 and 0.40, re-optimizes all the other parameters and recomputes the smoothed path. From 2007 on, when the earnings ratio is observed, the path moves by at most 0.3 percentage points. Before 2007 the path is identified with the information available for those years, the informality rate and the aggregates, while the model carries the earnings information back in time, and it moves by at most 1.4 points in the 1990s. Without the earnings observable nothing ties the path to a measured level, and the estimates without earnings are far from those with them (Section~\ref{sec:results}).

The earnings ratio therefore identifies the level and the trajectory of the informal share. The steady-state parameter $\theta$ is a nuisance parameter in this setting. The likelihood does not pin it down, its posterior combines the likelihood with the priors on the persistence of the shocks, and the path, which is what the paper measures, does not depend on it.

\begin{table}[H]
\centering
\caption{Informal share of output (\%) when the steady-state share is fixed and all other parameters are re-optimized, baseline model.}
\label{tab:invariance}
\footnotesize
\begin{tabular}{c c c c c c c}
\toprule
\textbf{Fixed $\theta$} & \textbf{1993Q2} & \textbf{2000Q4} & \textbf{2007Q2} & \textbf{2015Q2} & \textbf{2019Q2} & \textbf{2026Q2} \\
\midrule
0.10 & 30.1 & 29.9 & 26.7 & 25.9 & 23.2 & 19.9 \\
0.20 & 30.3 & 29.8 & 26.7 & 25.8 & 23.2 & 19.9 \\
0.30 & 31.4 & 30.3 & 26.8 & 25.8 & 23.2 & 20.1 \\
0.40 & 31.5 & 30.0 & 27.0 & 25.7 & 23.3 & 20.0 \\

\bottomrule
\end{tabular}
\end{table}

Section~\ref{sec:results} applies the observable to the data.

\section{Results}
\label{sec:results}

The baseline estimate uses the labor-only informal sector, the statutory non-wage costs of formal employees and all the data available from 1993Q2 to 2026Q2, namely the five aggregate growth rates, the level of the informality rate (including the ICML reconstruction of the labor market) and the value-added earnings ratio from 2007. We report the equal-technology estimate as an upper bound and the estimate without earnings as a point of reference.

\subsection{Posterior and convergence}

Table~\ref{tab:posterior} reports the priors, the posterior medians and 90\% intervals, the ratio of posterior to prior standard deviation and the convergence diagnostics. The largest rank-normalized $\hat R$ is 1.005, for $\theta$, and the smallest bulk and tail effective sample sizes are 1{,}718 and 1{,}067 out of 320{,}000 retained draws. Acceptance rates are between 0.23 and 0.30.

\begin{table}[H]
\centering
\caption{Prior and posterior distributions, baseline estimate (labor-only informal sector, statutory non-wage costs).}
\label{tab:posterior}
\scriptsize
\begin{tabular}{l l r c r r r}
\toprule
\textbf{Parameter} & \textbf{Prior (mean, sd)} & \textbf{Median} & \textbf{90\% interval} & \textbf{sd ratio} & $\hat R$ & \textbf{ESS} \\
\midrule
$\theta$ informal output share & Beta(0.35, 0.1) & 0.262 & [0.240, 0.277] & 0.12 & 1.005 & 1,718 \\
$\eta$ informal labor elasticity & Beta(0.7, 0.1) & 0.518 & [0.360, 0.680] & 0.97 & 1.001 & 4,464 \\
$\rho_\mu$ trend growth & Beta(0.4, 0.2) & 0.103 & [0.036, 0.186] & 0.23 & 1.004 & 2,190 \\
$\rho_g$ government & Beta(0.7, 0.15) & 0.816 & [0.737, 0.869] & 0.28 & 1.003 & 3,397 \\
$\rho_\nu$ investment efficiency & Beta(0.7, 0.15) & 0.977 & [0.898, 0.996] & 0.22 & 1.004 & 2,289 \\
$\rho_\psi$ labor supply & Beta(0.7, 0.15) & 0.993 & [0.981, 0.998] & 0.04 & 1.004 & 2,964 \\
$\rho_\omega$ formality wedge & Beta(0.85, 0.1) & 0.997 & [0.991, 0.999] & 0.03 & 1.003 & 3,645 \\
$\rho_x$ absorption & Beta(0.7, 0.15) & 0.895 & [0.751, 0.961] & 0.44 & 1.002 & 2,643 \\
$\sigma_\mu$ & Inv.\ gamma(0.01, 0.02) & 0.016 & [0.014, 0.019] & 0.07 & 1.001 & 4,868 \\
$\sigma_g$ & Inv.\ gamma(0.03, 0.06) & 0.052 & [0.047, 0.058] & 0.06 & 1.001 & 4,695 \\
$\sigma_\nu$ & Inv.\ gamma(0.05, 0.1) & 0.011 & [0.008, 0.015] & 0.02 & 1.002 & 2,873 \\
$\sigma_\psi$ & Inv.\ gamma(0.02, 0.04) & 0.020 & [0.017, 0.024] & 0.06 & 1.001 & 4,703 \\
$\sigma_\omega$ & Inv.\ gamma(0.02, 0.04) & 0.022 & [0.019, 0.027] & 0.06 & 1.002 & 4,249 \\
$\sigma_x$ & Inv.\ gamma(0.02, 0.04) & 0.007 & [0.006, 0.010] & 0.03 & 1.003 & 2,292 \\
$\sigma_{e,I}$ investment (level) & Inv.\ gamma(0.03, 0.03) & 0.042 & [0.026, 0.051] & 0.25 & 1.003 & 1,774 \\
$\sigma_{e,N}$ employment (level) & Inv.\ gamma(0.01, 0.01) & 0.010 & [0.009, 0.012] & 0.10 & 1.001 & 4,869 \\
$\sigma_{e,s}$ logit informality (level) & Inv.\ gamma(0.01, 0.01) & 0.011 & [0.007, 0.015] & 0.23 & 1.004 & 2,383 \\
$\rho_r$ informal productivity & Beta(0.8, 0.1) & 0.923 & [0.859, 0.962] & 0.32 & 1.003 & 3,234 \\
$\sigma_r$ & Inv.\ gamma(0.02, 0.04) & 0.016 & [0.011, 0.021] & 0.07 & 1.002 & 3,715 \\
$\sigma_{e,R}$ earnings ratio & Inv.\ gamma(0.03, 0.03) & 0.033 & [0.027, 0.040] & 0.13 & 1.002 & 4,353 \\
$b$ earnings shift from 2022Q1 & Normal(0, 0.2) & -0.035 & [-0.079, 0.009] & 0.13 & 1.002 & 4,142 \\

\bottomrule
\end{tabular}
\end{table}

\begin{figure}[H]
\centering
\includegraphics[width=0.75\textwidth]{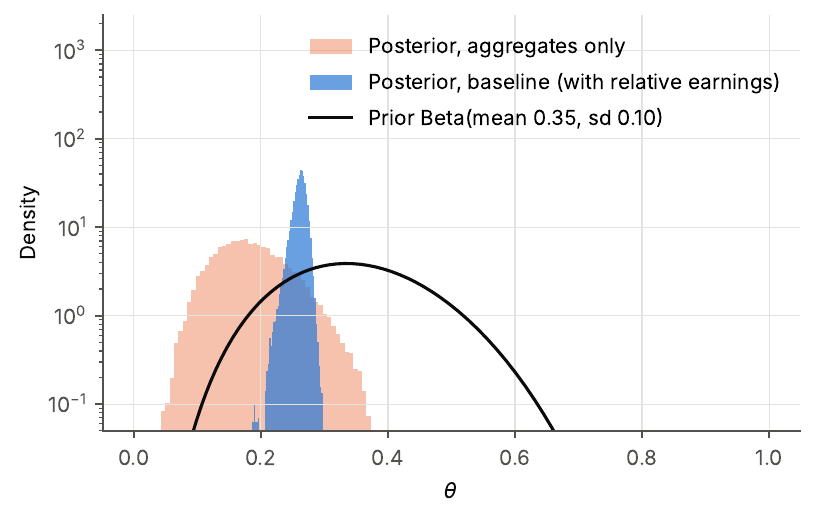}
\caption{Prior and posteriors of the steady-state informal share $\theta$: baseline with the earnings observable and estimate with aggregates only (logarithmic density scale).}
\label{fig:prior-posterior}
\end{figure}

The steady-state informal share has a posterior median of 0.262 and a 90\% interval of 0.240--0.277. Its posterior standard deviation is 12\% of the prior one, compared with 57\% when the earnings ratio is not observed (Figure~\ref{fig:prior-posterior}), although, as Section~\ref{sec:id-baseline} showed, part of this concentration comes from the priors on the persistence of the shocks, and what the data identify is the path of the share. The median earnings ratio is $R\approx0.31$, meaning that informal income per worker is about 31\% of the labor cost of a formal employee and informal value added per worker, $(1-\alpha)R$, about 20\% of formal value added per worker. The data are silent on the labor elasticity of informal production $\eta$, whose posterior equals its prior. The informal-productivity shock is persistent (0.92) and small (1.6\% a quarter), and the measurement error of the earnings ratio is 3.3\%. With formal employees in the denominator, the 2022 level shift of the earnings ratio is small and cannot be distinguished from zero ($-0.035$, with a 90\% interval from $-0.079$ to $0.009$).

\subsection{Fit}

Table~\ref{tab:fit} compares the model's root mean squared one-step-ahead prediction error with the standard deviation of each observable and with the error of a univariate first-order autoregression. Since quarterly growth rates have almost no persistence, the autoregression is the relevant benchmark. The model comes within about three percent of it for GDP, consumption and employment and does better for government consumption. The informality rate, observed in levels, is predicted with an error of 15\% of its standard deviation, close to that of the autoregression.

\begin{table}[H]
\centering
\caption{One-step-ahead prediction, baseline estimate (pandemic quarters excluded).}
\label{tab:fit}
\footnotesize
\begin{tabular}{l r r r r}
\toprule
\textbf{Observable} & \textbf{sd} & \textbf{Autocorrelation} & \textbf{RMSE/sd, model} & \textbf{RMSE/sd, AR(1)} \\
\midrule
GDP growth & 0.0150 & -0.13 & 1.014 & 0.992 \\
Private consumption growth & 0.0127 & 0.06 & 0.996 & 0.998 \\
Investment growth & 0.0844 & -0.45 & 0.900 & 0.892 \\
Government consumption growth & 0.0536 & -0.13 & 0.983 & 0.991 \\
Employment growth & 0.0148 & -0.21 & 1.013 & 0.979 \\
Logit informality rate (level) & 0.2008 & 0.99 & 0.153 & 0.122 \\

\bottomrule
\end{tabular}
\end{table}

\subsection{What drives informality}

Table~\ref{tab:variance} reports the forecast-error variance decomposition at the posterior median. The formality wedge explains most of the variance of the informality rate, from 67\% at one quarter to 92\% at twenty years, and informal productivity most of the rest. Unlike in the equal-technology model, aggregate shocks now move the informality rate (Proposition~\ref{prop:rate}), but only a little, with the trend shock accounting for about 4\% at short horizons and investment efficiency and labor supply for about 2\% at long horizons. The informal output share is driven by informal productivity at short horizons (73\% at one quarter) and by the wedge in the long run (66\%). Detrended output responds mainly to the labor-supply shock and to the wedge, which moves workers between sectors with very different value added per worker.

\begin{table}[H]
\centering
\caption{Forecast-error variance decomposition at the posterior median (\%), baseline estimate.}
\label{tab:variance}
\scriptsize
\begin{tabular}{l r rrrrrrr}
\toprule
\textbf{Variable} & \textbf{Horizon} & \textbf{Trend} & \textbf{Gov.} & \textbf{Inv.\ eff.} & \textbf{Labor supply} & \textbf{Wedge} & \textbf{Absorption} & \textbf{Inf.\ prod.} \\
\midrule
Detrended output & 1 & 6.8 & 0.6 & 5.0 & 38.0 & 33.8 & 1.4 & 14.4 \\
Detrended output & 4 & 6.0 & 0.3 & 5.7 & 39.3 & 35.6 & 0.9 & 12.1 \\
Detrended output & 20 & 3.3 & 0.1 & 8.1 & 41.7 & 41.0 & 0.2 & 5.6 \\
Detrended output & 80 & 1.1 & 0.0 & 6.9 & 40.8 & 49.3 & 0.1 & 1.8 \\
Logit informality rate & 1 & 3.7 & 0.0 & 0.0 & 0.3 & 67.0 & 0.0 & 29.0 \\
Logit informality rate & 4 & 3.9 & 0.0 & 0.1 & 0.4 & 71.6 & 0.0 & 23.9 \\
Logit informality rate & 20 & 2.5 & 0.1 & 1.2 & 1.0 & 84.9 & 0.1 & 10.3 \\
Logit informality rate & 80 & 0.8 & 0.0 & 1.9 & 2.0 & 92.2 & 0.0 & 3.0 \\
Logit informal output share & 1 & 9.2 & 0.0 & 0.1 & 0.7 & 16.8 & 0.0 & 73.2 \\
Logit informal output share & 4 & 10.9 & 0.0 & 0.4 & 1.0 & 21.0 & 0.0 & 66.6 \\
Logit informal output share & 20 & 9.6 & 0.2 & 4.6 & 4.0 & 41.8 & 0.2 & 39.6 \\
Logit informal output share & 80 & 3.7 & 0.1 & 8.3 & 8.8 & 65.5 & 0.2 & 13.4 \\

\bottomrule
\end{tabular}
\end{table}

\subsection{The informal share of output, 1993--2026}

\begin{figure}[H]
\centering
\includegraphics[width=0.9\textwidth]{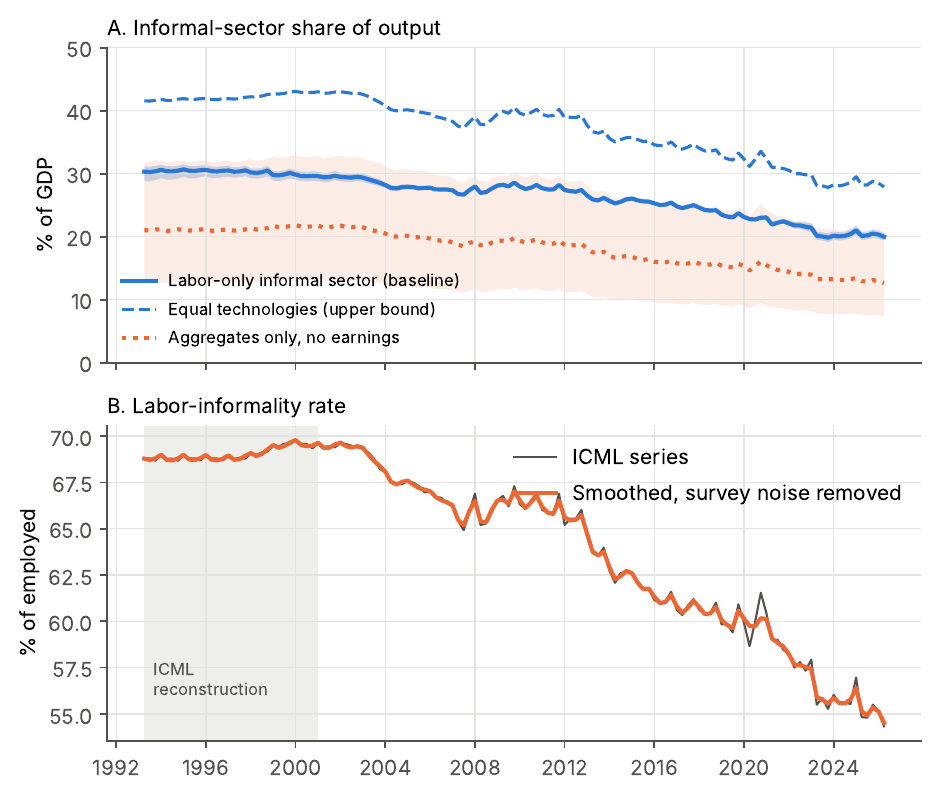}
\caption{Panel A: informal share of output, posterior median and 90\% band, baseline (labor-only informal sector, statutory non-wage costs), equal-technology estimate (upper bound) and estimate with aggregates only. Panel B: labor-informality rate from ICML and its smoothed value. The shaded area marks the ICML reconstruction (1993--2000).}
\label{fig:share}
\end{figure}

Figure~\ref{fig:share} shows the informal share of output computed from \eqref{eq:share} for 400 posterior draws, with the informality rate and the earnings ratio given by the Kalman smoother. The posterior median is 30.3\% of GDP in 1993 (90\% band in 1993Q2 28.8\%--31.2\%) and reaches its maximum, 30.7\%, in 1995. It is 29.7\% in 2000, 27.8\% in 2005 and 27.9\% in 2010, and then declines to 25.8\% in 2015, 23.4\% in 2019, 21.8\% in 2022 and 20.0\% in 2026Q2 (90\% band 19.4\%--20.7\%). The ten-point decline follows the fall of the informality rate from 69\% to 54\% of the employed, while the earnings ratio adds quarterly variation but no trend. The equal-technology estimate is eight to eleven points higher and the estimate without earnings seven to nine points lower.

\subsection{Other estimated series}

\begin{figure}[H]
\centering
\includegraphics[width=0.95\textwidth]{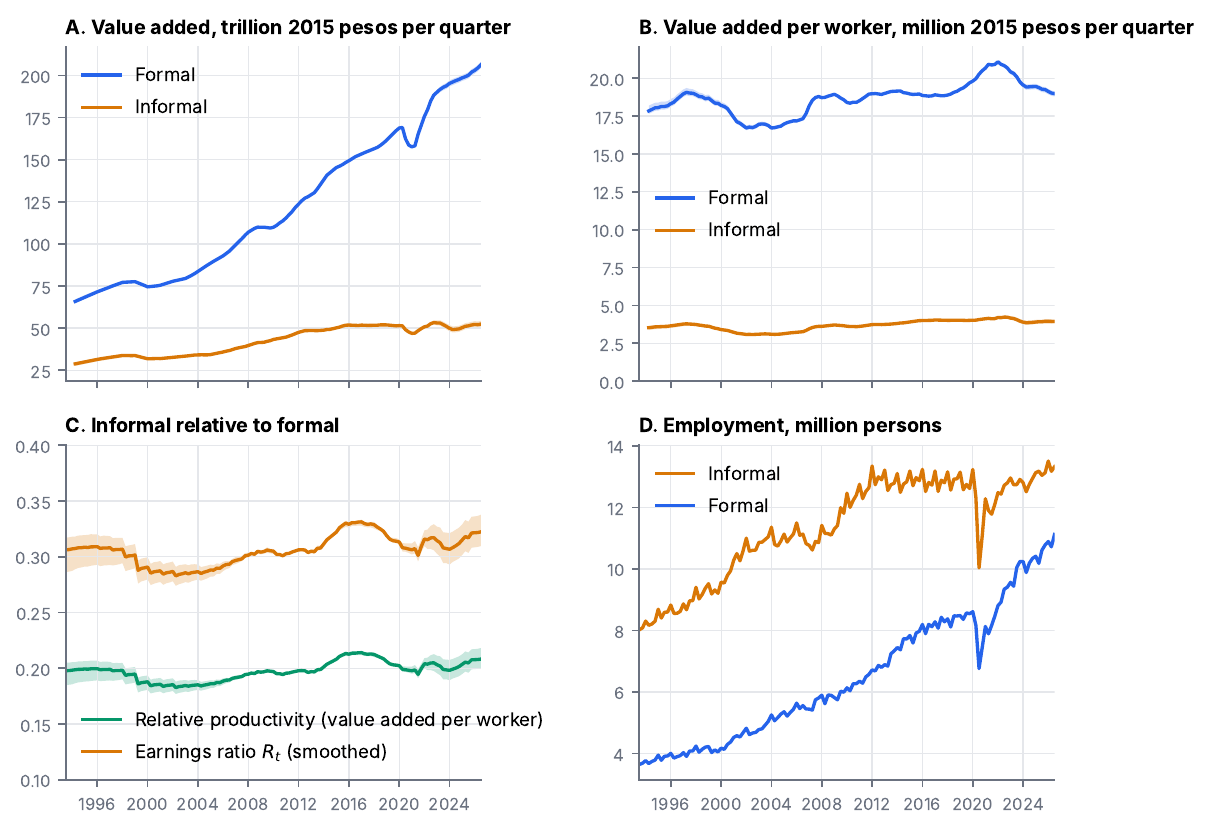}
\caption{Estimated series, baseline specification, posterior median and 90\% band (400 draws). Panels A and B: value added and value added per worker by sector, four-quarter moving averages (the national accounts are not seasonally adjusted). Panel C: informal-to-formal value added per worker and smoothed earnings ratio $R_t$. Panel D: informal and formal employment.}
\label{fig:series}
\end{figure}

The model estimates more than the informal share. Figure~\ref{fig:series} shows the other quarterly series implied by the smoothed states, and Table~\ref{tab:estimated-series} in Appendix~\ref{app:series} gives their annual values. These series and the seven structural shocks (Figure~\ref{fig:shocks}) are distributed with the paper at quarterly and annual frequency.

Informal value added grows from 120 trillion 2015 pesos in 1994 to 209 trillion in 2025, at 1.8\% a year, while formal value added grows at 3.6\%, from 274 to 812 trillion (panel A). The divergence comes from employment rather than from productivity. Informal employment rises from 8.2 to 13.3 million persons and formal employment triples, from 3.7 to 11.1 million (panel D). Value added per worker, by contrast, is roughly stable in both sectors: about 18--20 million 2015 pesos per quarter in the formal sector and 3.2--4.0 million in the informal sector (panel B). Informal value added per worker is therefore about one-fifth of formal value added per worker throughout the sample (panel C), between 0.18 in 2002 and 0.21 in 2016. It follows the smoothed earnings ratio $R_t$, scaled by the labor share $1-\alpha$, and falls during the 1999 crisis, recovers until 2016 and dips again during the pandemic. The informal share of output thus declines because the formal sector absorbs a growing share of workers, not because informal work becomes less productive.

\subsection{Comparison with published estimates}
\label{sec:benchmarks}

\subsubsection{Published series}

\begin{figure}[H]
\centering
\includegraphics[width=0.9\textwidth]{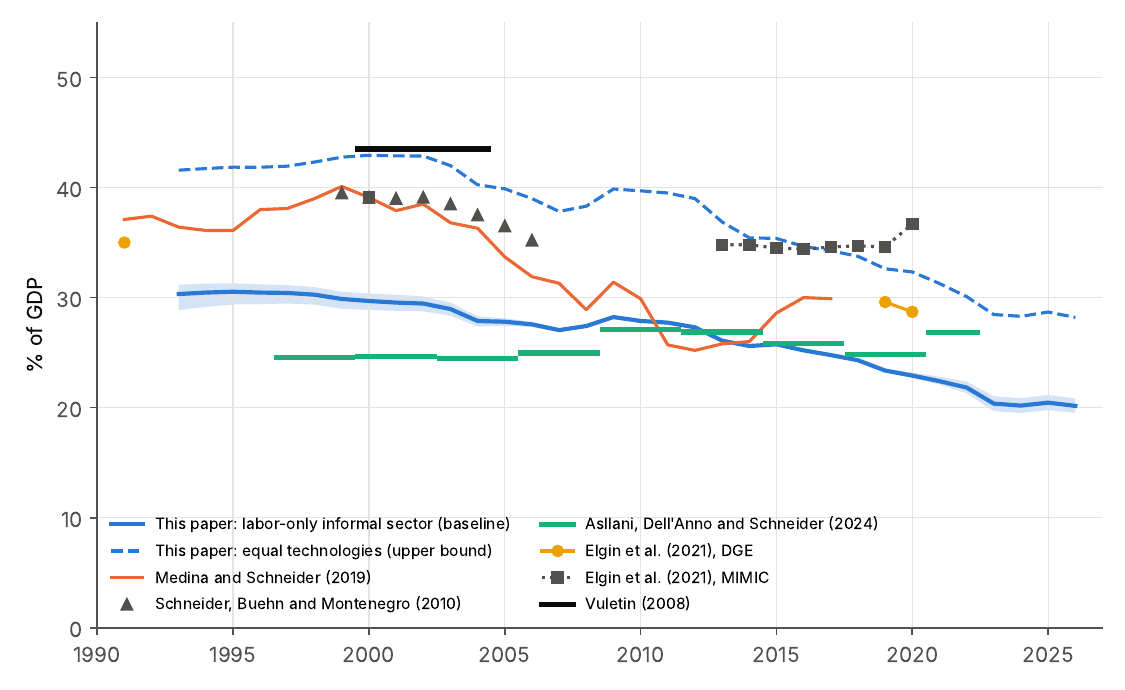}
\caption{Informal share of output: this paper and published estimates for Colombia (percent of GDP). Horizontal segments are period averages.}
\label{fig:benchmarks}
\end{figure}

Figure~\ref{fig:benchmarks} compares the estimate with the published annual series for Colombia in our reference collection, which fall into four families. The MIMIC estimates of Schneider and coauthors \citep{SchneiderBuehnMontenegro2010,MedinaSchneider2019} place the shadow economy at 36\%--40\% of official GDP in the 1990s and early 2000s and at 25\%--30\% after 2011. The enhanced MIMIC of \citet{AsllaniDellAnnoSchneider2024}, which restricts the concept to legal production deliberately concealed from the authorities, is almost flat at 24\%--27\%. The World Bank database \citep{ElginKoseOhnsorgeYu2021} reports a DGE series, built on the model of \citet{ElginOztunali2012}, that declines from 35\% in 1991 to 29\% in 2020, and a MIMIC series of 34\%--39\%. \citet{Vuletin2008} estimates 43.5\% for the early 2000s.

\subsubsection{How the estimate compares}

The published estimates differ from one another by up to 19 percentage points in the same year, and the baseline stays inside that range throughout. It is 6 to 10 points below the MIMIC series of Schneider and coauthors between 1993 and 2005 and within about three points of \citet{MedinaSchneider2019} in 2008--2015, when that series falls to 25\%--31\%. It is within about one point of \citet{AsllaniDellAnnoSchneider2024} for 2009--2020, two to six points above it before and five below in 2021--2022. It is about six points below the World Bank DGE series in 2019--2020 and nine to fourteen below its MIMIC series. Its decline is steeper than that of the MIMIC series, which change little by construction. The equal-technology estimate lies at the top of the published range, next to \citet{Vuletin2008} and the World Bank MIMIC series, and the estimate without earnings below all of them.

\subsubsection{Differences in concept}

The comparison has to be read with three conceptual differences in mind. MIMIC estimates are expressed as a percentage of official GDP and treat the shadow economy as lying outside it, whereas the Colombian national accounts already include an estimate of the non-observed economy and our share is a fraction of measured GDP. MIMIC levels are also not independent measurements, since they are converted into percentages of GDP with benchmark values taken from other studies. \citet{AsllaniDellAnnoSchneider2024}, for instance, anchor the scale to informal employment times an assumed informal wage, which amounts to the accounting of Proposition~\ref{prop:share} with the earnings ratio assumed rather than measured. Finally, our concept is the output of informal workers, which includes legal own-account production that is not concealed and excludes illegal activity. DANE's non-observed economy (22.4\% of gross production in the 2015 base) is not directly comparable either, because it refers to gross production rather than value added.

\subsection{Sensitivity}

The posterior bands describe parameter and state uncertainty given the model and the measurement equation, and the sensitivity analysis gauges the uncertainty about the measurement assumptions themselves. Table~\ref{tab:sensitivity} shows how the level changes with the concept of the formal wage (panel A), with other assumptions around the constant-wedge specification (panel B) and with the technology (panel C).

The way survey earnings are converted into labor cost matters. With contributions only, the 2026 share is 22.8\% instead of 20.0\%, and with a constant 25\% wedge it is 20.6\%. The capital share also matters, through the factor $1-\alpha$ in \eqref{eq:share}, and with $\alpha=0.50$ the 2026 share falls to 17.0\% from 20.6\%. By contrast, the prior of $\eta$, starting the sample in 2007 and observing the capital--output ratio in levels all leave the 2026 share within 0.2 points. Observing capital does raise the 1990s by about 3.5 points, since the model then has to accommodate the strong rise of the capital--output ratio over the sample. The equal-technology estimates bound the estimate from above.

\begin{table}[H]
\centering
\caption{Informal share of output under alternative specifications (posterior median and 90\% band, \%). $^{\dagger}$Largest $\hat R$ 1.053 (for $\theta$); the path is reported because it does not depend on $\theta$ (Table~\ref{tab:invariance}).}
\label{tab:sensitivity}
\scriptsize
\setlength{\tabcolsep}{4pt}
\begin{tabular}{l c c c c}
\toprule
\textbf{Specification} & $R$ & $\theta$ (steady state) & \textbf{1993Q2} & \textbf{2026Q2} \\
\midrule
\multicolumn{5}{l}{\emph{A. Formal wage concept (labor-only, $\alpha=0.354$, $\delta=4\%$)}} \\
Baseline: statutory non-wage costs, contributions and benefits & 0.31 & 0.262 [0.240, 0.277] & 30.3 [28.8, 31.2] & 20.0 [19.4, 20.7] \\
Statutory contributions only & 0.36 & 0.292 [0.251, 0.309] & 33.8 [31.9, 34.7] & 22.8 [22.1, 23.6] \\
Constant wedge, $\bar\tau=0.25$ & 0.34 & 0.277 [0.251, 0.293] & 32.7 [31.6, 33.7] & 20.6 [19.9, 21.3] \\
\multicolumn{5}{l}{\emph{B. Other assumptions (labor-only, constant wedge)}} \\
$\alpha=0.33$, $\delta=6\%$ & 0.34 & 0.284 [0.260, 0.299] & 33.4 [32.4, 34.4] & 21.1 [20.5, 21.8] \\
$\alpha=0.50$ & 0.34 & 0.235 [0.213, 0.253] & 27.6 [26.0, 28.9] & 17.0 [16.3, 17.7] \\
Prior of $\eta$ centered at 0.5 & 0.34 & 0.275 [0.247, 0.293] & 32.7 [31.7, 33.9] & 20.5 [19.9, 21.2] \\
Prior of $\eta$ centered at 0.9 & 0.34 & 0.281 [0.262, 0.294] & 32.7 [31.6, 33.6] & 20.6 [20.0, 21.4] \\
Sample from 2007Q1 & 0.34 & 0.256 [0.244, 0.266] & -- & 20.6 [20.0, 21.4] \\
Capital--output ratio observed in levels$^{\dagger}$ & 0.34 & 0.258 [0.166, 0.292] & 36.2 [35.7, 36.7] & 20.4 [19.8, 21.0] \\
\multicolumn{5}{l}{\emph{C. Equal technologies and aggregates only}} \\
Equal technologies, constant wedge (upper bound) & 0.33 & 0.371 [0.363, 0.377] & 41.6 [40.8, 42.2] & 27.9 [27.2, 28.5] \\
Equal technologies, no wedge adjustment & 0.44 & 0.440 [0.431, 0.447] & 48.7 [47.9, 49.3] & 34.0 [33.2, 34.6] \\
Equal technologies, hourly earnings & 0.35 & 0.389 [0.377, 0.397] & 43.4 [42.3, 44.1] & 29.4 [28.7, 30.0] \\
Aggregates only (no earnings) & 0.12 & 0.181 [0.101, 0.289] & 21.0 [12.8, 31.9] & 12.6 [7.4, 20.3] \\

\bottomrule
\end{tabular}
\end{table}

\begin{figure}[H]
\centering
\includegraphics[width=0.75\textwidth]{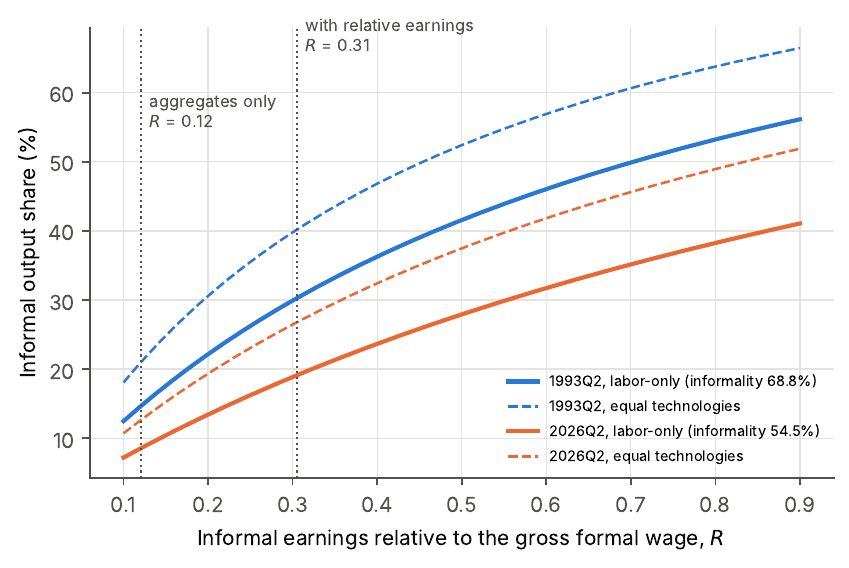}
\caption{Informal share of output implied by Proposition~\ref{prop:share} as a function of the earnings ratio $R$, at the informality rates of 1993Q2 and 2026Q2, for the labor-only and the equal-technology accounting. Dotted lines mark the posterior medians of $R$ with aggregates only and with the earnings observable.}
\label{fig:anchor}
\end{figure}

Every specification with earnings shows a decline of ten to fifteen points between 1993 and 2026. Figure~\ref{fig:anchor} places the estimates on the mapping of Proposition~\ref{prop:share}.

\section{Discussion: interpretation and implications}
\label{sec:discussion}

\subsection{What the estimate measures}

The estimate is the share of measured value added produced by informal work, as the household survey classifies it. Proposition~\ref{prop:share} ties it to three measurable quantities, the informality rate, relative earnings and the formal capital share, and the link uses the survey concepts directly. The income of own-account workers and informal employers is mixed income, which already includes the return to their own capital, and the income of informal employees is adjusted by group (Section~\ref{sec:earnings-data}). The posterior bands describe uncertainty given this mapping, and Table~\ref{tab:sensitivity} shows how the level moves under other mappings.

This concept is related to the shadow or non-observed economy but is not the same thing. Formal firms may hire workers informally, some own-account production is declared and included in the accounts, and illegal production falls outside the labor classification used here. The comparisons in Figure~\ref{fig:benchmarks} are therefore checks of magnitude and trend, and they are reassuring on both counts. The estimate lies within the range of the published series, very close to the enhanced MIMIC estimates in 2009--2020, and shares their downward trend.

\subsection{Identifying the path}

The identification exercises distinguish two objects. Aggregate data do not pin down the steady-state parameter $\theta$, which helps explain why calibrated and estimated DGE models in the literature reach very different levels for the same country. The quarterly path of the informal share, on the other hand, is identified once relative earnings are observed. From 2007 on it is determined by the observed informality rate, the observed earnings ratio and the capital share, and it varies by less than 0.3 percentage points across values of $\theta$. Before 2007 the path is identified with the information available for those years, the reconstructed informality rate and the national accounts, and the model carries the earnings information back; this step moves the 1990s by at most 1.4 points across values of $\theta$ between 0.10 and 0.40. Simulations with known parameters recover the true share with high precision.

For this literature, the practical lesson is that the level of the informal economy in a DGE model should be identified with a direct measure of relative productivity, rather than through the aggregates or by calibration, and that identification checks should be reported together with the posterior.

\subsection{Implications for research and policy}

The contribution of informal work to output has fallen together with labor informality, from about 30\% to 20\% of GDP over three decades. The estimates also measure the productivity gap behind the difference between the two, since at the end of the sample informal value added per worker is about one-fifth of the formal level. That gap is the relevant magnitude for policy. Formalizing a worker does not raise output by itself, and what matters is whether the worker moves into more productive activity. Evaluating specific reforms, such as changes in payroll taxes or enforcement, is a natural next step for the model, which already contains the formality wedge through which such policies operate.

Because the data and code are public and the earnings series comes from the regular household survey, the estimate can be updated every quarter, which makes it a timely complement to DANE's periodic measurement of the non-observed economy.

\section{Scope and extensions}
\label{sec:limitations}

The estimates rest on choices made with the information available. All of them are explicit and can be tested, and each suggests an extension that would sharpen the results.

\subsection{From earnings to value added}

The baseline treats the mixed income of own-account workers and employers as their value added and adjusts the earnings of informal employees by group. The sensitivity analysis shows how much the level depends on this treatment, and the equal-technology specification bounds it from above. Direct data on the value added and capital of informal units, such as DANE's micro-business survey (EMICRON), would allow the mapping to be estimated instead of assumed.

\subsection{Labor costs and the capital share}

Formal labor costs are computed from the statutory non-wage costs, and actual payments may differ because of exemptions or noncompliance; with contributions only, the 2026 share would be 22.8\% instead of 20.0\%. The capital share of 0.354 is estimated from Colombian data and agrees with the GRECO historical accounts (0.36--0.42), while a value of 0.50 would lower the 2026 share to 17\%. Firm-level data on labor costs and sectoral estimates of the capital share would narrow this range.

\subsection{The historical period}

Before 2007 there are no consolidated microdata on informal earnings, and the national-accounts and labor-market series for the 1990s are reconstructions. The paper uses all the information that exists for those years, and the invariance and sensitivity results show how much the early path depends on the model. Historical survey microdata, if they are ever recovered, could enter the estimation directly, since the Kalman filter already handles missing observations.

\subsection{Model structure}

A single labor-only informal technology stands in for a heterogeneous sector of own-account workers, employees of unregistered units and informally hired workers in registered firms. Splitting the informal sector by type of worker, and bringing observed policy instruments into the formality wedge, would make the model usable for policy evaluation.

None of these choices changes the main results, which are the accounting relationship, the identification of the path through relative earnings and the decline of the informal share from about 30\% to 20\% of GDP.

\section{Conclusions}
\label{sec:conclusions}

This paper estimates the contribution of informal work to Colombian output from 1993 to 2026. In a two-sector DSGE model, the informal output share depends on three quantities, the proportion of workers employed informally, their value added per worker relative to formal workers and the capital share of formal production. The first is observed in the informality rate, the second is measured with a new quarterly series of relative earnings built from household-survey microdata, and the third is estimated from Colombian data.

The identification analysis shows why the relative-earnings series is needed. Likelihood profiles, prior--posterior comparisons and simulations with known parameters all indicate that national accounts and labor-market aggregates do not determine the size of the informal sector on their own. Once relative earnings are observed, the quarterly path of the informal share is identified, does not depend on the model's steady state and is recovered precisely in simulations. Distinguishing the path, which the data identify, from the steady-state parameter, which they need not identify, is what makes it possible to estimate the informal economy with a DSGE model.

Under the baseline specification, the informal share falls from about 30\% of GDP in 1993 to 20\% in 2026Q2, while labor informality falls from about 69\% to 54\%. Informal value added per worker is about one-fifth of formal value added per worker, which explains how a majority of workers can be informal while producing a fifth of output. The estimates lie within the range of published Colombian series and close to the most recent MIMIC estimates.

The model also delivers quarterly informal and formal GDP, relative productivity, the informality rate net of measurement error and the structural shocks behind them, all of which are published with the paper. The data and code allow the series to be updated every quarter and extended as new information becomes available, in particular direct measures of the value added of informal units and observed policy instruments that would turn the model into a tool for evaluating formalization policies.

\section*{Ethical Statement}

This study uses aggregated official statistics, a public research dataset derived from them, and the anonymized public-use microdata of DANE's household survey (GEIH), from which only quarterly aggregates are computed. The analysis uses no directly identifying survey records and involves no contact with research participants. Ethics-review requirements, if any, remain subject to the author's institutional rules.

\section*{Acknowledgments}

The author acknowledges the work of the Departamento Administrativo Nacional de Estad\'istica (DANE) in producing and publishing the statistics on which this study relies. The views and conclusions expressed here are solely those of the author. The results are research estimates and are not official statistics.

\section*{Funding and Competing Interests}

This research received no funding from public agencies, private institutions or nonprofit organizations. The author declares no competing interests.

\section*{Data and Code Availability}

The labor series are those of ICML release 2.6.1, archived at \url{https://doi.org/10.5281/zenodo.24357243}. The GEIH microdata are published by DANE in its microdata catalog. The quarterly dataset used in the estimation, the relative-earnings series, the script that computes it from the GEIH microdata, the estimation code (model, state space, sampler, diagnostics and the earnings measurement equation), the result files and the script that regenerates the figures and tables are distributed as ancillary files of this paper. They are also maintained at \url{https://github.com/jaimevera1107/informalidad_dsge_colombia}, where a single command downloads the new national accounts, ICML and GEIH releases and updates the estimated series every quarter.

\appendix
\section{The model in detail}
\label{app:model}

This appendix lists the equations exactly as they are solved and estimated. The code that implements them is in the ancillary files (\code{m2/model.py} and \code{m2/statespace.py}).

\subsection{Stationary equilibrium conditions}

Let $c_t=C_t/X_t$, $y_t=Y_t/X_t$, $y_{f,t}=Y_{f,t}/X_t$, $y_{i,t}=Y_{i,t}/X_t$, $i_t=I_t/X_t$, $g_t=G_t/X_t$, and $k_t=K_t/X_{t-1}$. Capital per unit of the current trend is $\tilde k_t = K_t/X_t = k_t\,e^{-(\gamma+\mu_t)}$. The thirteen equations, in the order of the code, are:
\begin{align}
&\text{(1) Output:} && y_t = y_{f,t} + y_{i,t},\quad y_{f,t}=\tilde k_t^{\alpha}N_{f,t}^{1-\alpha},\quad y_{i,t}=A_i e^{z^r_t}N_{i,t}^{\eta} \\
&\text{(2) Formal labor:} && \psi_t\chi_fN_{f,t}^{\varphi}c_t = (1-\bar\tau)e^{-\omega_t}(1-\alpha)\,y_{f,t}/N_{f,t} \\
&\text{(3) Informal labor:} && \psi_t\chi_iN_{i,t}^{\varphi}c_t = \eta\,y_{i,t}/N_{i,t} \\
&\text{(4) Resources:} && y_t(1+\bar x + x_t) = c_t + i_t + g_t,\quad g_t=\bar g\,e^{z^g_t} \\
&\text{(5) Capital:} && k_{t+1} = (1-\delta)\tilde k_t + \nu_t i_t \\
&\text{(6) Euler:} && \frac{1}{\nu_tc_t} = \beta E_t\Big[\frac{e^{-(\gamma+\mu_{t+1})}}{c_{t+1}}\Big(\alpha\frac{y_{f,t+1}}{\tilde k_{t+1}} + \frac{1-\delta}{\nu_{t+1}}\Big)\Big] \\
&\text{(7)--(13) Shocks:} && \zeta_{t+1}=\rho_\zeta\zeta_t+\sigma_\zeta\varepsilon_{\zeta,t+1},\quad \zeta\in\{\mu, z^g,\ln\nu,\ln\psi,\omega,x,z^r\}
\end{align}
The factor $e^{-(\gamma+\mu_{t+1})}$ in the Euler equation appears because $C_{t+1}=c_{t+1}X_{t+1}$ and $X_{t+1}=X_te^{\gamma+\mu_{t+1}}$: a positive trend shock raises future consumption and lowers its marginal utility. In equation (5), capital installed in $t$ is divided by $X_t$ and becomes $k_{t+1}$.

\subsection{Steady state}

At $\zeta_t=0$ for every shock and with the calibration of Table~\ref{tab:calibration}:
\begin{align*}
r &= e^{\gamma}/\beta-1+\delta, & \tilde k/y_f &= \alpha/r, & N_f &= \bar N(1-\bar s),\quad N_i=\bar N\bar s,\\
y_f &= (\tilde k/y_f)^{\alpha/(1-\alpha)}N_f, & y &= y_f/(1-\theta), & y_i&=\theta y,\quad A_i = y_i/N_i^{\eta},\\
k &= e^{\gamma}\tilde k, & i &= k-(1-\delta)\tilde k, & c &= (C/Y)\,y,\quad g=(G/Y)\,y,\\
\bar x &= (c+i+g)/y-1, & \chi_f &= \frac{(1-\bar\tau)(1-\alpha)y_f/N_f}{N_f^{\varphi}c}, & \chi_i &= \frac{\eta\,y_i/N_i}{N_i^{\varphi}c}.
\end{align*}
The steady-state earnings ratio that enters the measurement equation is
\[
\bar R = \frac{y_i/N_i}{(1-\alpha)\,y_f/N_f} = \frac{\theta}{(1-\theta)(1-\alpha)}\cdot\frac{1-\bar s}{\bar s},
\]
which is Proposition~\ref{prop:share} inverted at the steady state. At the posterior median ($\theta=0.277$, $\alpha=0.354$, $\bar s=0.645$) it is 0.33.

\subsection{Linearization and solution}

Let $\bm{z}_t$ be the vector of the thirteen variables in logarithms (shocks in levels) and $\bar{\bm{z}}$ its steady state. Writing the system as $\bm{f}(\bm{z}_t, E_t\bm{z}_{t+1})=\bm{0}$, the first-order approximation is
\[
\bm{F}_1\,(\bm{z}_t-\bar{\bm{z}}) + \bm{F}_2\,E_t(\bm{z}_{t+1}-\bar{\bm{z}}) = \bm{0},
\]
where $\bm{F}_1$ and $\bm{F}_2$ are the Jacobians with respect to current and expected future values, computed by central finite differences. The generalized Schur (QZ) decomposition of the pencil $(-\bm{F}_1,\bm{F}_2)$ is ordered so that stable eigenvalues come first \citep{Klein2000}. With eight stable eigenvalues for eight states the solution is unique and takes the form \eqref{eq:solution}; otherwise the parameter draw is rejected. The solution is checked by verifying that $\bm{F}_1\bm{M} + \bm{F}_2\bm{M}\bm{T} = \bm{0}$, where $\bm{M}$ maps the states into all variables.

\subsection{Measurement equations}

Hats denote log deviations from the steady state and $q(t)$ the quarter of the year. The observables of the baseline are linked to the model as follows.
\begin{align*}
\Delta\ln Y^{\mathrm{obs}}_t - m^{Y}_{q(t)} &= \hat y_t - \hat y_{t-1} + \mu_t + u^Y_t,\\
\Delta\ln C^{\mathrm{obs}}_t - m^{C}_{q(t)} &= \hat c_t - \hat c_{t-1} + \mu_t + u^C_t,\\
\Delta\ln I^{\mathrm{obs}}_t - m^{I}_{q(t)} &= \hat\imath_t - \hat\imath_{t-1} + \mu_t + e^I_t - e^I_{t-1} + u^I_t,\\
\Delta\ln G^{\mathrm{obs}}_t - m^{G}_{q(t)} &= z^g_t - z^g_{t-1} + \mu_t + u^G_t,\\
\Delta\ln N^{\mathrm{obs}}_t - m^{N}_{q(t)} &= (1-\bar s)(\hat n_{f,t}-\hat n_{f,t-1}) + \bar s(\hat n_{i,t}-\hat n_{i,t-1}) + e^N_t - e^N_{t-1} + u^N_t,\\
\logit s^{\mathrm{obs}}_t - \logit\bar s - \kappa_{q(t)} &= \hat n_{i,t} - \hat n_{f,t} + e^s_t,\\
\ln R^{\mathrm{obs}}_t - \ln\bar R
&= \ln(1+d_t^{\mathrm{nw}}) + b\,\mathbb{1}_{t\ge2022Q1}
  + z^r_t + (\eta-1)\hat n_{i,t} \\
&\quad - \alpha(\hat k_t - \mu_t - \hat n_{f,t}) + e^R_t .
\end{align*}
The trend shock $\mu_t$ appears in the growth rates because detrended variables are measured relative to $X_t$, whose growth is $\gamma+\mu_t$; the mean growth $\gamma$ is removed with the quarter-of-year means $m_q$. Total employment is linearized as a weighted average of the two types of work. The errors $e^I$, $e^N$ are level errors carried as states, $e^s$ and $e^R$ are independent errors of the two observables in levels, and $u$ is the small growth-rate error that keeps the covariance of the observables non-singular. $\kappa_q$ is the quarter-of-year pattern of the logit informality rate over 2005--2018, centered at zero. In the last equation, $\ln(1+d_t^{\mathrm{nw}})$ converts the wages reported in the survey into the employer's labor cost.

\subsection{Likelihood and sampler}

Stacking the states of \eqref{eq:solution}, their lags and the level errors in a vector $\bm{\xi}_t$, the model and the measurement equations form a linear Gaussian state-space system,
\[
\bm{\xi}_{t+1} = \bm{A}\bm{\xi}_t + \bm{B}\bm{\varepsilon}_{t+1},\qquad \bm{y}^{\mathrm{obs}}_t = \bm{Z}\bm{\xi}_t + \bm{u}_t .
\]
The Kalman filter computes, quarter by quarter, the forecast of the observables given the past, the forecast error $\bm{v}_t$ and its covariance $\bm{F}_t$, and the log likelihood is $-\tfrac12\sum_t\big(\ln|2\pi\bm{F}_t| + \bm{v}_t^{\top}\bm{F}_t^{-1}\bm{v}_t\big)$ over the observed entries. Quarters with missing observations use only the entries that are observed. The posterior is the likelihood times the prior; the random-walk Metropolis--Hastings sampler proposes a new parameter vector near the current one and accepts it with probability $\min\{1, p(\text{new})/p(\text{current})\}$, so that after a burn-in the accepted draws are a sample from the posterior. The Kalman smoother then recovers, for each draw, the path of the states given all the data, from which the informal share of output \eqref{eq:share} is computed.

\section{Estimated series}
\label{app:series}

\begin{figure}[H]
\centering
\includegraphics[width=0.95\textwidth]{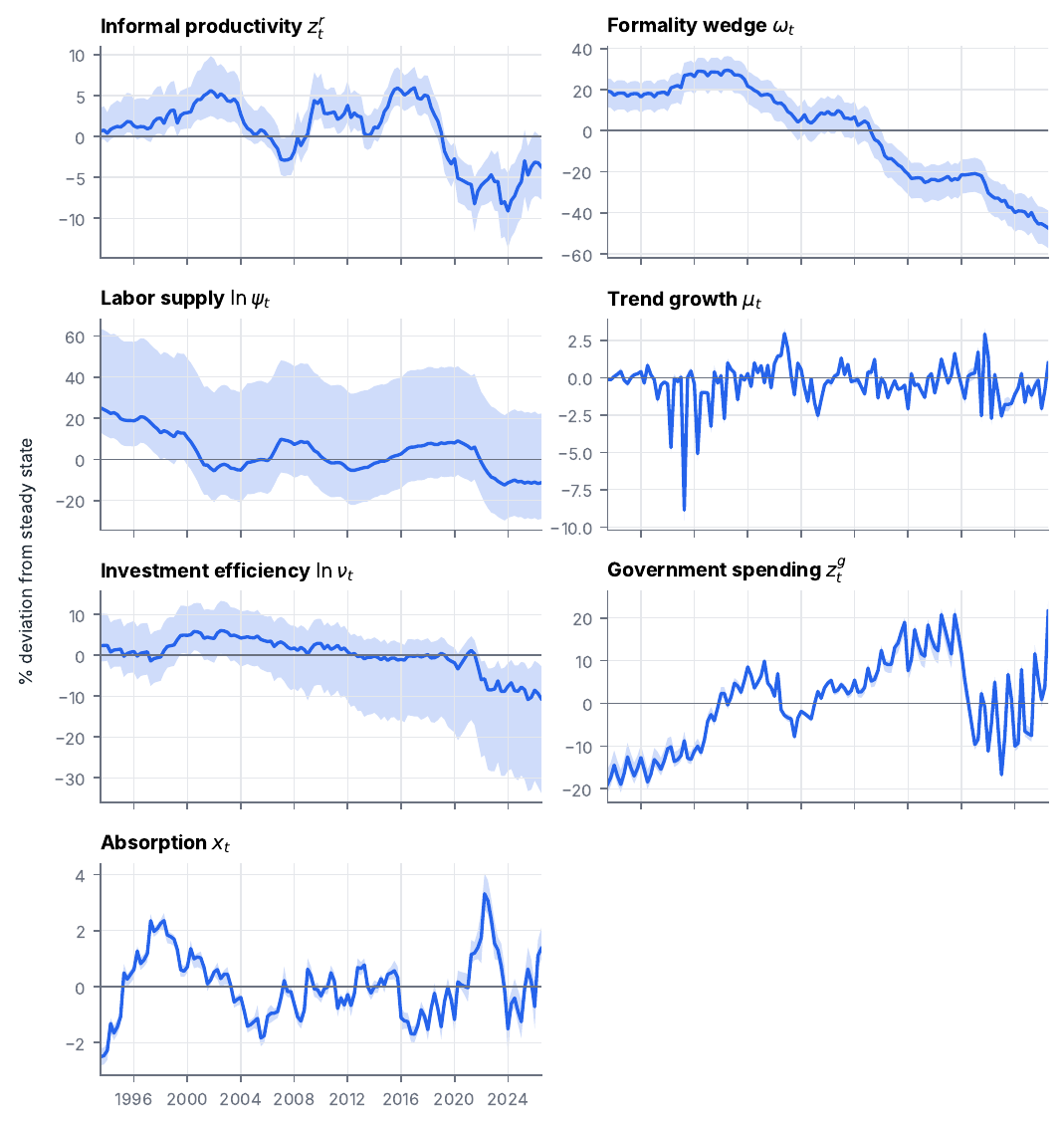}
\caption{Smoothed structural shocks, baseline specification, posterior median and 90\% band, percent deviation from the steady state.}
\label{fig:shocks}
\end{figure}

Table~\ref{tab:estimated-series} reports annual values of the series estimated by the baseline specification; quarterly values and the full set of percentiles are in the ancillary files. Annual value added is the sum of the four quarters, and the annual informal share is annual informal value added over annual GDP. Figure~\ref{fig:shocks} shows the smoothed structural shocks.

\begin{table}[H]
\centering
\caption{Annual estimated series, baseline specification. Value added in trillion 2015 pesos; posterior medians and, for the informal share, 90\% bands (400 draws). Relative productivity is informal value added per worker over formal value added per worker.}
\label{tab:estimated-series}
\footnotesize
\setlength{\tabcolsep}{5pt}
\begin{tabular}{rrrrrrrr}
\toprule
 & GDP & \multicolumn{2}{c}{Informal share of output (\%)} & Informal VA & Formal VA & Labor & Relative \\
Year & (trillion) & Median & 90\% band & (trillion) & (trillion) & informality (\%) & productivity \\
\midrule
1994 & 393.4 & 30.5 & [29.2, 31.2] & 120.0 & 273.5 & 68.8 & 0.199 \\
1995 & 411.9 & 30.6 & [29.4, 31.3] & 125.9 & 286.0 & 68.8 & 0.199 \\
1996 & 428.4 & 30.5 & [29.4, 31.2] & 130.5 & 297.8 & 68.8 & 0.199 \\
1997 & 443.4 & 30.4 & [29.5, 31.1] & 134.9 & 308.5 & 68.8 & 0.198 \\
1998 & 445.1 & 30.3 & [29.4, 30.9] & 134.7 & 310.3 & 69.1 & 0.194 \\
1999 & 425.0 & 29.9 & [29.1, 30.5] & 127.1 & 298.0 & 69.5 & 0.187 \\
2000 & 429.2 & 29.7 & [29.0, 30.3] & 127.5 & 301.6 & 69.6 & 0.185 \\
2001 & 441.9 & 29.6 & [28.9, 30.2] & 130.8 & 311.2 & 69.5 & 0.185 \\
2002 & 450.7 & 29.5 & [28.9, 30.1] & 132.9 & 317.8 & 69.5 & 0.184 \\
2003 & 471.3 & 29.0 & [28.4, 29.5] & 136.5 & 334.8 & 68.9 & 0.185 \\
2004 & 491.3 & 27.9 & [27.4, 28.3] & 137.0 & 354.3 & 67.6 & 0.185 \\
2005 & 514.9 & 27.8 & [27.5, 28.1] & 143.3 & 371.6 & 67.3 & 0.187 \\
2006 & 549.4 & 27.6 & [27.4, 27.8] & 151.5 & 397.9 & 66.7 & 0.190 \\
2007 & 586.5 & 27.0 & [26.9, 27.2] & 158.6 & 427.9 & 65.7 & 0.193 \\
2008 & 605.7 & 27.4 & [27.3, 27.5] & 166.0 & 439.7 & 65.8 & 0.196 \\
2009 & 612.6 & 28.2 & [28.1, 28.4] & 173.0 & 439.6 & 66.6 & 0.197 \\
2010 & 640.2 & 27.9 & [27.8, 27.9] & 178.5 & 461.7 & 66.5 & 0.195 \\
2011 & 684.6 & 27.7 & [27.7, 27.8] & 189.9 & 494.7 & 66.1 & 0.197 \\
2012 & 711.4 & 27.3 & [27.2, 27.4] & 194.2 & 517.2 & 65.6 & 0.197 \\
2013 & 747.9 & 26.1 & [26.0, 26.2] & 195.0 & 552.9 & 63.9 & 0.199 \\
2014 & 781.6 & 25.6 & [25.5, 25.7] & 200.1 & 581.4 & 62.6 & 0.206 \\
2015 & 804.7 & 25.7 & [25.6, 25.9] & 207.2 & 597.5 & 62.1 & 0.212 \\
2016 & 821.5 & 25.2 & [25.0, 25.3] & 207.1 & 614.4 & 61.2 & 0.214 \\
2017 & 832.7 & 24.8 & [24.6, 24.9] & 206.4 & 626.3 & 60.8 & 0.213 \\
2018 & 854.0 & 24.3 & [24.2, 24.4] & 207.6 & 646.5 & 60.6 & 0.209 \\
2019 & 881.2 & 23.4 & [23.3, 23.5] & 205.9 & 675.3 & 60.0 & 0.203 \\
2020 & 817.9 & 22.9 & [22.7, 23.2] & 187.3 & 630.6 & 59.9 & 0.198 \\
2021 & 906.2 & 22.4 & [22.1, 22.8] & 203.0 & 703.2 & 59.2 & 0.199 \\
2022 & 972.7 & 21.8 & [21.2, 22.4] & 212.2 & 760.5 & 57.8 & 0.204 \\
2023 & 980.8 & 20.3 & [19.7, 21.0] & 199.4 & 781.4 & 56.2 & 0.199 \\
2024 & 995.5 & 20.2 & [19.5, 20.9] & 200.9 & 794.6 & 55.7 & 0.201 \\
2025 & 1,021.1 & 20.4 & [19.8, 21.2] & 208.7 & 812.3 & 55.5 & 0.206 \\
\bottomrule
\end{tabular}

\end{table}

\begin{thebibliography}{33}
\providecommand{\natexlab}[1]{#1}
\providecommand{\url}[1]{\texttt{#1}}
\expandafter\ifx\csname urlstyle\endcsname\relax
  \providecommand{\doi}[1]{doi: #1}\else
  \providecommand{\doi}{doi: \begingroup \urlstyle{rm}\Url}\fi

\bibitem[Aguiar and Gopinath(2007)]{AguiarGopinath2007}
Mark Aguiar and Gita Gopinath.
\newblock Emerging market business cycles: The cycle is the trend.
\newblock \emph{Journal of Political Economy}, 115\penalty0 (1):\penalty0
  69--102, 2007.
\newblock \doi{10.1086/511283}.

\bibitem[An and Schorfheide(2007)]{AnSchorfheide2007}
Sungbae An and Frank Schorfheide.
\newblock {Bayesian} analysis of {DSGE} models.
\newblock \emph{Econometric Reviews}, 26\penalty0 (2--4):\penalty0 113--172,
  2007.
\newblock \doi{10.1080/07474930701220071}.

\bibitem[Arango et~al.(2006)Arango, Misas, and L{\'o}pez]{ArangoMisasLopez2006}
Carlos~A. Arango, Martha~A. Misas, and Enrique L{\'o}pez.
\newblock Econom{\'i}a subterr{\'a}nea en {Colombia} 1976--2003: una
  medici{\'o}n a partir de la demanda de efectivo.
\newblock \emph{Ensayos sobre Pol{\'i}tica Econ{\'o}mica}, 24\penalty0 (50),
  2006.

\bibitem[Asllani et~al.(2024)Asllani, Dell'Anno, and
  Schneider]{AsllaniDellAnnoSchneider2024}
Alban Asllani, Roberto Dell'Anno, and Friedrich Schneider.
\newblock Long-run estimates of the global informal economies and new insights
  for 152 countries over 1997 to 2022 using an enhanced {MIMIC} approach.
\newblock IZA Discussion Paper 17557, IZA Institute of Labor Economics, 2024.

\bibitem[Barrios et~al.(1993)Barrios, Henao, Posada, Valderrama, and
  V{\'a}squez]{Barrios1993}
Adriana Barrios, Mar{\'i}a~L. Henao, Carlos~E. Posada, F.~Valderrama, and
  Diego~M. V{\'a}squez.
\newblock Empleo y capital en {C}olombia: nuevas estimaciones (1950--1992).
\newblock Archivos de macroeconom{\'i}a, Departamento Nacional de
  Planeaci{\'o}n, 1993.

\bibitem[Busato and Chiarini(2004)]{BusatoChiarini2004}
Francesco Busato and Bruno Chiarini.
\newblock Market and underground activities in a two-sector dynamic equilibrium
  model.
\newblock \emph{Economic Theory}, 23\penalty0 (4):\penalty0 831--861, 2004.

\bibitem[Canova and Sala(2009)]{CanovaSala2009}
Fabio Canova and Luca Sala.
\newblock Back to square one: Identification issues in {DSGE} models.
\newblock \emph{Journal of Monetary Economics}, 56\penalty0 (4):\penalty0
  431--449, 2009.

\bibitem[{DANE}(2025)]{DANE_IMIE}
{DANE}.
\newblock {\'I}ndice multidimensional de informalidad empresarial ({IMIE})
  2024.
\newblock Departamento Administrativo Nacional de Estad{\'i}stica, press
  release, 2025.
\newblock URL
  \url{https://www.dane.gov.co/files/operaciones/IMIE/cp-IMIE-2024.pdf}.

\bibitem[Denton(1971)]{Denton1971}
Frank~T. Denton.
\newblock Adjustment of monthly or quarterly series to annual totals: An
  approach based on quadratic minimization.
\newblock \emph{Journal of the American Statistical Association}, 66\penalty0
  (333):\penalty0 99--102, 1971.
\newblock \doi{10.1080/01621459.1971.10482227}.

\bibitem[{Departamento Administrativo Nacional de Estad{\'i}stica
  (DANE)}(2021)]{DANE_SIECI}
{Departamento Administrativo Nacional de Estad{\'i}stica (DANE)}.
\newblock Plan general del sistema de informaci{\'o}n para actividades
  informales ({SIECI}).
\newblock
  \url{https://www.dane.gov.co/files/proyectos-de-resoluciones-y-decretos/p-2021-resolucion-SIECI/Plan_general_SIECI.pdf},
  2021.

\bibitem[Durbin and Koopman(2012)]{DurbinKoopman2012}
James Durbin and Siem~Jan Koopman.
\newblock \emph{Time Series Analysis by State Space Methods}.
\newblock Oxford University Press, 2 edition, 2012.

\bibitem[Elgin and {\"O}ztunal{\i}(2012)]{ElginOztunali2012}
Ceyhun Elgin and O{\u{g}}uz {\"O}ztunal{\i}.
\newblock Shadow economies around the world: Model based estimates.
\newblock Working Paper~5, Bogazici University, Department of Economics, 2012.

\bibitem[Elgin et~al.(2021)Elgin, Kose, Ohnsorge, and
  Yu]{ElginKoseOhnsorgeYu2021}
Ceyhun Elgin, M.~Ayhan Kose, Franziska Ohnsorge, and Shu Yu.
\newblock Understanding informality.
\newblock CEPR Discussion Paper 16497, Centre for Economic Policy Research,
  2021.

\bibitem[Fern{\'a}ndez and Meza(2015)]{FernandezMeza2015}
Andr{\'e}s Fern{\'a}ndez and Felipe Meza.
\newblock Informal employment and business cycles in emerging economies: The
  case of {Mexico}.
\newblock \emph{Review of Economic Dynamics}, 18\penalty0 (2):\penalty0
  381--405, 2015.

\bibitem[{GRECO}(1999)]{GRECO1999}
{GRECO}.
\newblock El crecimiento econ{\'o}mico colombiano en el siglo {XX}: aspectos
  globales.
\newblock Borradores de Econom{\'i}a 134, Banco de la Rep{\'u}blica, 1999.

\bibitem[Ihrig and Moe(2004)]{IhrigMoe2004}
Jane Ihrig and Karine~S. Moe.
\newblock Lurking in the shadows: The informal sector and government policy.
\newblock \emph{Journal of Development Economics}, 73\penalty0 (2):\penalty0
  541--557, 2004.

\bibitem[Iskrev(2010)]{Iskrev2010}
Nikolay Iskrev.
\newblock Local identification in {DSGE} models.
\newblock \emph{Journal of Monetary Economics}, 57\penalty0 (2):\penalty0
  189--202, 2010.

\bibitem[Klein(2000)]{Klein2000}
Paul Klein.
\newblock Using the generalized {Schur} form to solve a multivariate linear
  rational expectations model.
\newblock \emph{Journal of Economic Dynamics and Control}, 24\penalty0
  (10):\penalty0 1405--1423, 2000.
\newblock \doi{10.1016/S0165-1889(99)00045-7}.

\bibitem[Koop et~al.(2013)Koop, Pesaran, and Smith]{KoopPesaranSmith2013}
Gary Koop, M.~Hashem Pesaran, and Ron~P. Smith.
\newblock On identification of {Bayesian} {DSGE} models.
\newblock \emph{Journal of Business \& Economic Statistics}, 31\penalty0
  (3):\penalty0 300--314, 2013.

\bibitem[Medina and Schneider(2018)]{MedinaSchneider2018}
Leandro Medina and Friedrich Schneider.
\newblock Shadow economies around the world: What did we learn over the last 20
  years?
\newblock IMF Working Paper 18/17, International Monetary Fund, 2018.

\bibitem[Medina and Schneider(2019)]{MedinaSchneider2019}
Leandro Medina and Friedrich Schneider.
\newblock Shedding light on the shadow economy: A global database and the
  interaction with the official one.
\newblock CESifo Working Paper 7981, CESifo, 2019.

\bibitem[Orsi et~al.(2014)Orsi, Raggi, and Turino]{OrsiRaggiTurino2014}
Renzo Orsi, Davide Raggi, and Francesco Turino.
\newblock Size, trend, and policy implications of the underground economy.
\newblock \emph{Review of Economic Dynamics}, 17\penalty0 (3):\penalty0
  417--436, 2014.
\newblock \doi{10.1016/j.red.2013.11.001}.

\bibitem[Restrepo-Echavarria(2014)]{RestrepoEchavarria2014}
Paulina Restrepo-Echavarria.
\newblock Macroeconomic volatility: The role of the informal economy.
\newblock \emph{European Economic Review}, 70:\penalty0 454--469, 2014.

\bibitem[Roberts et~al.(1997)Roberts, Gelman, and
  Gilks]{RobertsGelmanGilks1997}
Gareth~O. Roberts, Andrew Gelman, and Walter~R. Gilks.
\newblock Weak convergence and optimal scaling of random walk {Metropolis}
  algorithms.
\newblock \emph{The Annals of Applied Probability}, 7\penalty0 (1):\penalty0
  110--120, 1997.
\newblock \doi{10.1214/aoap/1034625254}.

\bibitem[Rodr{\'i}guez~Mu{\~n}oz et~al.(2004)Rodr{\'i}guez~Mu{\~n}oz,
  Perilla~Jim{\'e}nez, and Reyes~Pe{\~n}a]{RodriguezPerillaReyes2004}
Jorge~Iv{\'a}n Rodr{\'i}guez~Mu{\~n}oz, Juan~Ricardo Perilla~Jim{\'e}nez, and
  Jos{\'e}~Daniel Reyes~Pe{\~n}a.
\newblock C{\'a}lculo del {PIB} potencial en {C}olombia: 1970--2003.
\newblock Archivos de Econom{\'i}a 261, Departamento Nacional de
  Planeaci{\'o}n, 2004.

\bibitem[Schneider and Enste(2000)]{SchneiderEnste2000}
Friedrich Schneider and Dominik~H. Enste.
\newblock Shadow economies: Size, causes, and consequences.
\newblock \emph{Journal of Economic Literature}, 38\penalty0 (1):\penalty0
  77--114, 2000.
\newblock \doi{10.1257/jel.38.1.77}.

\bibitem[Schneider et~al.(2010)Schneider, Buehn, and
  Montenegro]{SchneiderBuehnMontenegro2010}
Friedrich Schneider, Andreas Buehn, and Claudio~E. Montenegro.
\newblock New estimates for the shadow economies all over the world.
\newblock \emph{International Economic Journal}, 24\penalty0 (4):\penalty0
  443--461, 2010.
\newblock \doi{10.1080/10168737.2010.525974}.

\bibitem[Solis-Garcia and Xie(2017)]{SolisGarciaXie2017}
Mario Solis-Garcia and Yingtong Xie.
\newblock Measuring the size of the shadow economy using a dynamic general
  equilibrium model with trends.
\newblock MPRA Paper 78968, Munich Personal RePEc Archive, 2017.
\newblock URL \url{https://mpra.ub.uni-muenchen.de/78968/}.

\bibitem[Stock and Watson(1993)]{StockWatson1993}
James~H. Stock and Mark~W. Watson.
\newblock A simple estimator of cointegrating vectors in higher order
  integrated systems.
\newblock \emph{Econometrica}, 61\penalty0 (4):\penalty0 783--820, 1993.

\bibitem[Vehtari et~al.(2021)Vehtari, Gelman, Simpson, Carpenter, and
  B{\"u}rkner]{Vehtari2021}
Aki Vehtari, Andrew Gelman, Daniel Simpson, Bob Carpenter, and Paul-Christian
  B{\"u}rkner.
\newblock Rank-normalization, folding, and localization: An improved
  $\widehat{R}$ for assessing convergence of {MCMC}.
\newblock \emph{Bayesian Analysis}, 16\penalty0 (2):\penalty0 667--718, 2021.
\newblock \doi{10.1214/20-BA1221}.

\bibitem[Vera-Jaramillo(2025)]{VeraJaramillo2025ICML}
Jaime~A. Vera-Jaramillo.
\newblock Reconstructing subnational labor indicators in {Colombia}: An
  integrated machine and deep learning approach.
\newblock arXiv:2508.12514, 2025.

\bibitem[Vera-Jaramillo(2026)]{VeraJaramillo2026Zenodo}
Jaime~A. Vera-Jaramillo.
\newblock Reconstructing subnational labor indicators in {Colombia}
  (1993--2026): Datasets, release 2.6.1, 2026.

\bibitem[Vuletin(2008)]{Vuletin2008}
Guillermo Vuletin.
\newblock Measuring the informal economy in {L}atin {A}merica and the
  {C}aribbean.
\newblock IMF Working Paper 08/102, International Monetary Fund, 2008.

\end{thebibliography}
\end{document}